\documentclass[10 pt,journal]{IEEEtran}
\IEEEoverridecommandlockouts
\usepackage{amsmath,amssymb}
\usepackage{graphicx}
\usepackage{cite}
\usepackage{epstopdf}
\usepackage{color}
\usepackage{balance}
\usepackage{algorithm,float}
\usepackage{algorithmicx}
\usepackage{algpseudocode}
\usepackage{subfigure}

\ifCLASSINFOpdf
\else
\fi

	\newcommand{\algoname}[1]{\textnormal{\textsc{#1}}}

\begin{document}
	
	%
	\title{Coordinated Path Following Control of Fixed-wing Unmanned Aerial Vehicles}
	%
	%
	%
	\author{Hao~Chen, Yirui~Cong,~Xiangke~Wang,~Xin~Xu, and~Lincheng~Shen
		\thanks{			
			
			The authors are with the College
			of Intelligence Science and Technology, National University of Defense Technology, Changsha 410073,
			China. (e-mail: xkwang@nudt.edu.cn, xinxu@nudt.edu.cn).}}

	%
	%

	%
	\maketitle
	
{\color{red}This work has been submitted to the IEEE for possible publication. Copyright may be transferred without notice, after which this version may no longer be accessible.}

\begin{abstract}
In this paper, we investigate the problem of coordinated path following for fixed-wing UAVs with speed constraints in 2D plane. The objective is to steer a fleet of UAVs along the path(s) while achieving the desired sequenced inter-UAV arc distance. In contrast to the previous coordinated path following studies, we are able through our proposed hybrid control law to deal with the forward speed and the angular speed constraints of fixed-wing UAVs. More specifically, the hybrid control law makes all the UAVs work at two different levels: those UAVs whose path following errors are within an invariant set (i.e., the designed coordination set) work at the coordination level; and the other UAVs work at the single-agent level. At the coordination level, we prove that even with speed constraints, the proposed control law can make sure the path following errors reduce to zero, while the desired arc distances converge to the desired value. At the single-agent level, the convergence analysis for the path following error entering the coordination set is provided. We develop a hardware-in-the-loop simulation testbed of the multi-UAV system by using actual autopilots and the X-Plane simulator. The effectiveness of the proposed approach is corroborated with both MATLAB and the testbed.
\end{abstract}
	

	%
	\IEEEpeerreviewmaketitle
	
	\newtheorem{assumption}{Assumption}
	\newtheorem{definition}{Definition}
	\newtheorem{lemma}{Lemma}
	\newtheorem{theorem}{Theorem}
	\newtheorem{proposition}{Proposition}
	\newtheorem{corollary}{Corollary}
	\newtheorem{remark}{Remark}
	\newtheorem{problem}{Problem}
	\newtheorem{statement}{Statement}
	\newtheorem{example}{Example}

	\section{Introduction}\label{sec:Introduction}
	
	\subsection{Motivation}\label{sec:Motivation}
	
	Coordinated path following control of multiple fixed-wing unmanned aerial vehicles (UAVs) has attracted significant attention in recent years, due to its increasing demands in civil and military uses~\cite{Chung2018}.
	It studies how to steer a group of fixed-wing UAVs moving along given/planned paths while forming a desired formation pattern based on local interactions.

	Different from multirotor UAVs or ground vehicles, a fixed-wing UAV cannot move backwards and is actually constrained by a minimum forward speed which generates sufficiently large lift to support the UAV in flight.
	It should be noted that with the minimum forward speed constraint, the coordinated path following control behaves very differently:
	\begin{itemize}
		\item   During path following, the maximum angular speed is not negligible any more, since it combined with the minimum forward speed determines the minimum turning radius.
		As a result, one UAV can only follow the class of paths (curves) with limited curvature, which is determined by the minimum forward speed and the maximum angular speed;
		while the existing studies in the coordinated path following control problems did not take the speed-related curvature into account.
		\item   During the formation pattern forming process, one UAV cannot completely stop or become unacceptably slow to wait for another UAV.
		Consequently, some coordinating UAVs can quit the group formation, if the minimum forward speed and the maximum angular speed are not properly considered when designing the control law.\footnote{We have proved that the conventional coordinated path following control laws in~\cite{YingLan} cannot solve the coordinated path following control problem for fixed-wing UAVs if the constraints are not properly considered.}
	\end{itemize}

	Therefore, the coordinated path following control of fixed-wing UAVs is very different from that without minimum forward speed and maximum angular speed constraints in the literature, and cannot be solved by the existing methods (e.g.,~\cite{YingLan}).
	It is necessary to design an efficient control law to solve the coordinated path following control problem for fixed-wing UAVs.
	%


	%
	%
	%
	%

	\subsection{Related Work}\label{sec:Related Work}

The coordinated path following control problem consists of two core components: path following and multi-vehicle coordination.

	Path following problem has a long history dating back to the end of last century~\cite{Samson}. 
	For fixed-wing UAV path following, some early work mainly focused on the straight line following and circle following~\cite{Nelson2007Vector,Park2007Performance,Ambrosino2009Path,Sanghyo2010}, and a comparison of existing methods dealing with these two kinds of path following problem can be found in~\cite{Sujit2014}.
	In recent years, general curved path following problem has received increasing attention, and typical approaches are largely based on proportional-integral-derivative (PID) controller~\cite{rhee2010tight}, vector field method~\cite{Liang2016,Shulong2018}, sliding mode controller~\cite{Jackson2008}, backstepping controller~\cite{Ahmed2010}, adaptive controller~\cite{CaoStabilization}, nested saturation and control Lyapunov function based method~\cite{Shulong2017}, etc.
	
	In terms of the multi-vehicle coordination, typical methods include leader-follower approach~\cite{Roldao2014,hung2017q,Jin2019}, virtual structure approach~\cite{Norman2008,Ghommam2010,Li2012}, behavior-based approach~\cite{Antonelli2010,Werfel2014}.  We recommend the surveys in~\cite{Chung2018,Oh2015,WangXiangke2016} to readers who are interested in the existing coordinated control results.
	Some methods have been validated by the formation flight of fixed-wing UAVs: in~\cite{Wilson2016}, a vision-aided close formation flight of two UAVs was conducted based on leader-follower control;
	the work in~\cite{Zhiyong2019} considered several formation design approaches, and performed field experiments of two UAVs with a combined controller;
	as a big advance in terms of the scale, autonomous flight of 50 fixed-wing platforms was presented in~\cite{Chung2016}, but details of their control law were not provided.

	To combine the path following controller with the coordinated control architecture is the key factor to achieve coordinated path following. There have been a large number of studies dealing with the coordinated path following control of multirotor UAVs~\cite{Kushleyev2013,Roldao2014,Cichella2016Safe} and ground vehicles~\cite{Ghommam2010,Reyes2015}.
	However, most of these results cannot be applied to fixed-wing UAVs, since they are constrained by the minimum forward speed, which is a hard constraint in this kind of vehicles for certain missions.
	Note that the speed constraints have been partially considered in the coordinated path following problems for fixed-wing UAVs: in~\cite{Gonalves}, a centralized strategy was investigated for a group of fixed-wing UAVs to follow closed intersecting curves, where the coordination refers to the collision avoidance among the UAVs;
	%
	 in~\cite{Xargay2013}, a time-critical coordinated path following control problem was studied, and the proposed algorithm can steer a fleet of fixed-wing UAVs along given paths while arriving at their own final destinations at the same time;
	%
	%
	the cooperative moving path following problem was introduced in~\cite{Yuanzhe2019}, with sufficient conditions derived under which the closed-loop system is asymptotically stable. Nevertheless, in these studies, the forward speed constraints and the angular speed constraints are not considered as a whole.

	\subsection{Our Contributions}\label{sec:Our Contributions}
	
	In this work, we focus on the coordinated path following control problem for a group of fixed-wing UAVs.
	%
	To solve this problem, we propose a novel distributed hybrid control law by extending  the hybrid controller provided in~\cite{YingLan}. It should be noted that most of the existing work use Lyapunov analysis to conclude the stability of the hybrid systems~\cite{liberzon2003switching}. However, in this paper,  the stability of  the overall system is concluded by employing the similar technique with~\cite{YingLan}, which can provide a complete dynamical analysis of the overall systems.
	%
	%
	The main contributions of this paper are listed as follows:

	\begin{itemize}
		\item We propose a hybrid control framework based on an invariant set, the \emph{coordination set}, to solve the coordinated path following problem of fixed-wing UAVs.
		Different from the existing results, the coordination set is designed to guarantee the convergence of the path following error and coordination error while satisfying both the forward speed constraints and the angular speed constraints.
		%
		%
		\item We propose a coordinated path following control law inside the coordination set, and theoretically prove that even with the speed constraints of fixed-wing UAVs, the proposed control law can make the path following errors reduce to zero, while the desired arc distances converge to the desired value.
		\item We propose a single-agent level control law outside the coordination set by using optimal control, and the convergence analysis for the UAVs entering the coordination set is provided.
		\item We develop a hardware-in-the-loop simulation testbed of the multi-UAV system by using actual autopilots and the X-Plane simulator, and validate the proposed coordinated path following approach with the testbed.
	\end{itemize}
	
	\subsection{Paper Organization}\label{sec:Paper Organization}
	
	The paper is organized as follows. Section~\ref{sec:statements} presents the system model as well as the problem description, 
	and analyzes the difficulty from speed constraints.
Then, in the next two sections, we present our control law to tackle the coordinated path following problem with speed constraints: at coordination level when the path following error is within a coordination set (see Section~\ref{sec:S1}); and at single-agent level when it is outside this coordination set (see Section~\ref{sec:S2}).
In Section~\ref{sec:simulation} we show the simulation results, including the simulation with MATLAB and the hardware-in-the-loop simulation with the X-Plane simulator; and finally the concluding remarks are given in Section~\ref{sec:conclusion}.

	\subsection{Notation}\label{sec:Notation}
	
	A curve is said to be $\mathcal{C}^k$-smooth, if it admits an analytic expression $\Gamma(x,y)$, whose $k$\textsuperscript{th} derivative exists and is continuous. 
	The curvature of a path at any point $p$ on the curve is denoted as $\kappa (p)$. $\mathbf{T}(p)$ denotes the unit tangent vector of the curve at point $p$, and the curvature of the curve at point $p$ is defined as $\kappa (p)={\mathrm{d}\mathbf{T}(p)}/{\mathrm{d}s}$, where $s$ is the natural parameter of the curve representing the length of the curve. 
	%
	
	
	\section{Coordinated Path Following Problem for Fixed-wing UAVs}\label{sec:statements}
	
	In this section, we formulate the 2D coordinated path following problem for a group of $n$ homogeneous fixed-wing UAVs. The aim is to design a control law such that each UAV follows a predefined curved path while the sequenced inter-UAV arc distances converge to the desired constant.

	\subsection{System Model and Problem Description}
	
	Consider a group of fixed-wing UAVs flying 
	at the same altitude.
	Then, the state of the $i$\textsuperscript{th} UAV can be represented by the configuration vector $q_i=(x_i,y_i,\theta_i)^T\in\mathbb{R}^2\times[-\pi,\pi)$, where $(x_i,y_i)$ is the $i$\textsuperscript{th} UAV's position defined in an inertia coordinate frame $\mathcal{W}$, and $\theta_i$ is the orientation of the $i$\textsuperscript{th} UAV with respect to the $x$-axis of $\mathcal{W}$. The kinematic model of the $i$\textsuperscript{th} UAV with pure rolling and non-slipping is given as
	\begin{equation}
	\label{eq1}
	\dot{q_i}=  \begin{bmatrix}
	\dot{x_i}\\
	\dot{y_i}\\
	\dot{\theta_i}\\
	\end{bmatrix}
	=   \begin{bmatrix}
	\cos\theta_i & 0\\
	\sin\theta_i & 0\\
	0 & 1\\
	\end{bmatrix}
	\begin{bmatrix}
	v_i\\
	\omega_i\\
	\end{bmatrix},
	\end{equation}
	where control inputs $v_i$ and $\omega_i$ stand for the forward speed and angular speed of the $i$\textsuperscript{th} UAV, respectively.

	For fixed-wing UAVs, the speed constraints must be considered in~\eqref{eq1}.
	On the one hand, the forward speed of a UAV is constrained with saturation and dead zone, i.e., each fixed-wing UAV has the maximum and minimum forward speed constraints.
	On the other hand, the angular speed is also constrained with saturation.
	Mathematically, for the $i$\textsuperscript{th} UAV, the speed constraints are:
	\begin{equation}
	\label{eq_constraint}
	\begin{cases}
	0<v_{\min}\leq v_i\leq v_{\max},\\
	|\omega_i|\leq \omega_{\max},
	\end{cases}
	\end{equation}
	where $v_{\min}$ and $v_{\max}$ are the minimum and maximum forward speeds, while $\omega_{\max}$ is the maximum angular speed.
		We consider all the UAVs follow a directed curved path $\Gamma \in \mathcal{C}^2$.

		\begin{assumption}
			\label{asp_path}
			$\Gamma$ 
			is globally known to each UAV, and the absolute value of the curvature of $\Gamma$ at any point $p$ is less than a constant $\kappa_0$, i.e., $|\kappa(p)|<\kappa_0$.
		\end{assumption}
	%

\begin{figure}[!htbp]
		\centering
		\includegraphics[scale=0.35]{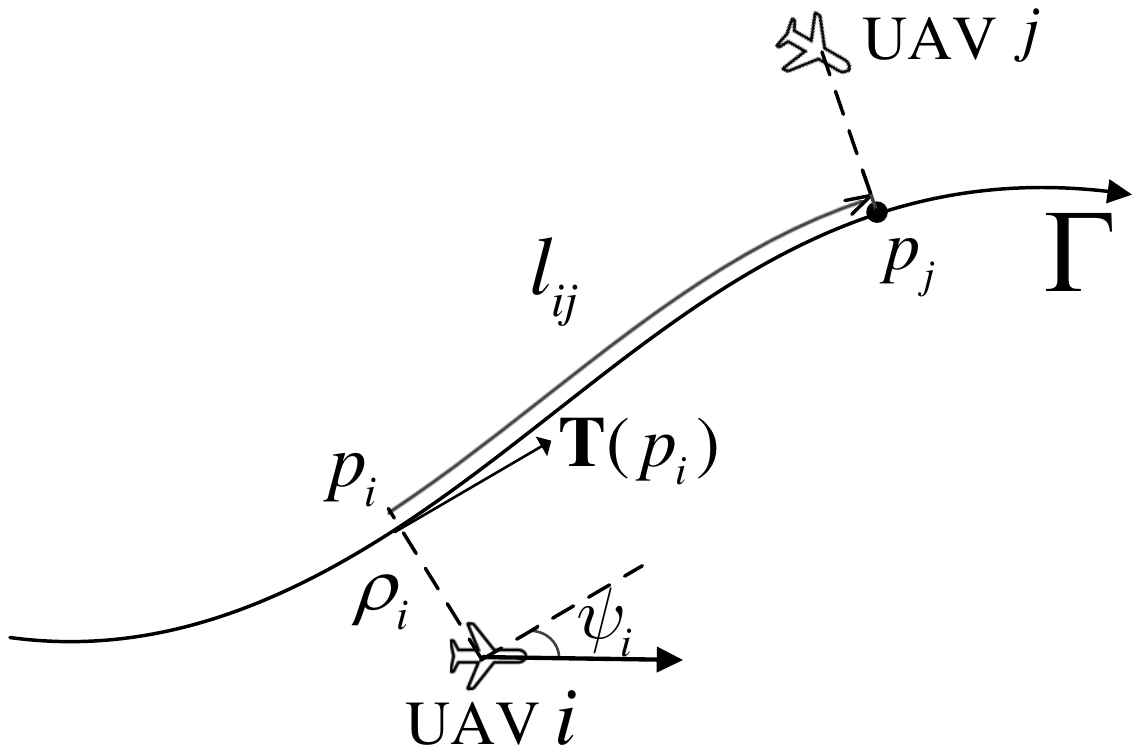}
		\caption{Path $\Gamma$ with a direction.
			Points $p_i$ and $p_j$ are the projections of the $i$\textsuperscript{th} and the $j$\textsuperscript{th} UAVs, respectively.
			$\phi_i=(\rho_i,\psi_i)$ is the path following error of the $i$\textsuperscript{th} UAV, and arc distance $l_{ij}$ is the length along the path from $p_i$ to $p_j$.
			In this case, the $j$\textsuperscript{th} UAV is the pre-neighbor of the $i$\textsuperscript{th} UAV.}
		\label{fig1}
	\end{figure}

	We note that due to the constraints on the forward and angular speeds, $\kappa_0$ must be not larger than $\omega_{\max}/v_{\min}$.
	This implies each UAV has a minimum turning radius (numerically equals $v_{\min}/\omega_{\max}$) when following a path.
	We use $\mathbf{T}(p)$ to denote the tangent vector of the path $\Gamma$ at point $p$.
		The point $p=(p_x,p_y)^T$ on the path is said to be a projection of the $i$\textsuperscript{th} UAV, if the vector $\boldsymbol{t}=(x_i-p_x,y_i-p_y)^T$ is orthogonal to $\mathbf{T}(p)$.	
	We define $\phi_i=(\rho_i,\psi_i)$ as the path following error of the $i$\textsuperscript{th} UAV with respect to the path $\Gamma$ (see Fig.~\ref{fig1}), where $\rho_i\in\mathbb{R}$ is the distance from the $i$\textsuperscript{th} UAV to its closest projection $p_i$ on $\Gamma$ with a sign:
	$\rho_i>0$ when the $i$\textsuperscript{th} UAV is on the left side of $\Gamma$ in the direction of the path, and $\rho_i<0$ when it is on the other side.
	For example, in Fig.~\ref{fig1}, we have $\rho_i<0$ for the $i$\textsuperscript{th} UAV, and $\rho_j>0$ for the $j$\textsuperscript{th} UAV.
	$\psi_i$ is the heading of the $i$\textsuperscript{th} UAV with respect to $\mathbf{T}(p_i)$ at its projection $p_i$.
	In this way, $\rho_i$ can be seen as the \emph{location difference} of the $i$\textsuperscript{th} UAV with respect to the path $\Gamma$; and $\psi_i\in[-\pi,\pi)$ is the \emph{orientation difference} between the heading of the $i$\textsuperscript{th} UAV and $\mathbf{T}(p_i)$. 
	
	\begin{remark}
		If $|\rho_i|<R_0$, where $R_0=\frac{1}{\kappa_0}$, the closest projection $p_i$ is unique~\cite{Samson}.
	\end{remark}
	
	The dynamics~\eqref{eq1} thus can be rewritten in the form of $\dot\phi_i=f(\phi_i)$
	%
	with the speed constraints represented by~\eqref{eq_constraint}:
	\begin{equation}
	\label{eq_patherror}
	\begin{cases}
	\dot{\rho_i}=v_i\sin\psi_i,\\
	\dot{\psi_i}=\omega_i-\frac{\kappa(p_i)v_i\cos\psi_i}{1-\kappa(p_i)\rho_i},
	\end{cases} i=1,...,n.
	\end{equation}
	UAV $i$ asymptotically follows the path $\Gamma$ if and only if $\phi_i \to \mathbf{0}$.

	In our coordinated path following problem, it not only requires all the UAVs move along the path $\Gamma$, but also guarantees the distance between any two adjacent UAVs is a desired constant, say $L$, in the sense of arc length.
	To describe adjacent UAVs, we introduce the definition of \emph{pre-neighbor} as follows.

	\begin{definition}\label{def:Pre-neighbor}
		The $j$\textsuperscript{th} UAV is the $i$\textsuperscript{th} UAV's \emph{pre-neighbor} if
		\begin{itemize}
			\item[i)] $|\rho_i|<R_0$, $|\rho_j|<R_0$;
			\item[ii)] The projection point $p_j$ of the $j$\textsuperscript{th} UAV on the path $\Gamma$ is in front of the $i$\textsuperscript{th} UAV's projection point $p_i$ (see Fig.~\ref{fig1}), without any other UAV's projection in the middle.
		\end{itemize}
	\end{definition}
	

		The two conditions in Definition~\ref{def:Pre-neighbor} imply that:
		(i) the pre-neighbor definition only works for the UAVs close enough to the path $\Gamma$ (i.e., with $|\rho_i|<R_0$); and
		(ii) each UAV can at most have one pre-neighbor.

\begin{remark}
When there are two or more UAVs with the same projection point on $\Gamma$, we define the pre-neighbor in ascending order with respect to the UAVs' labels to avoid ambiguity.

\end{remark}

	Since the desired inter-UAV arc distance is between adjacent UAVs, the $i$\textsuperscript{th} UAV just needs to focus on the arc distance $l_{ij}$ to its pre-neighbor.
	To simplify the description, we denote the arc distance between the $i$\textsuperscript{th} UAV and its pre-neighbor as $\zeta_i$.
	%
	%
	The coordination error for the $i$\textsuperscript{th} UAV is thus $L-\zeta_i$.

	Now we can formally define the coordinated path following problem for fixed-wing UAVs as follows.

	\begin{problem}[Coordinated Path Following]\label{prob:Coordinated Path Following}
		Given $n$ fixed-wing UAVs, each is modeled as~\eqref{eq1} with speed constraints represented by~\eqref{eq_constraint}, design control law such that $\phi_i \to \mathbf{0}$, and $L-\zeta_i \to 0$, $i=1,\ldots, n$,
		%
	\end{problem}

	
	%
	%

	\subsection{The Challenge from Speed Constraints}\label{sub:a}
	
\begin{figure}[!htb]  	
		\centering
		\includegraphics[scale=0.25]{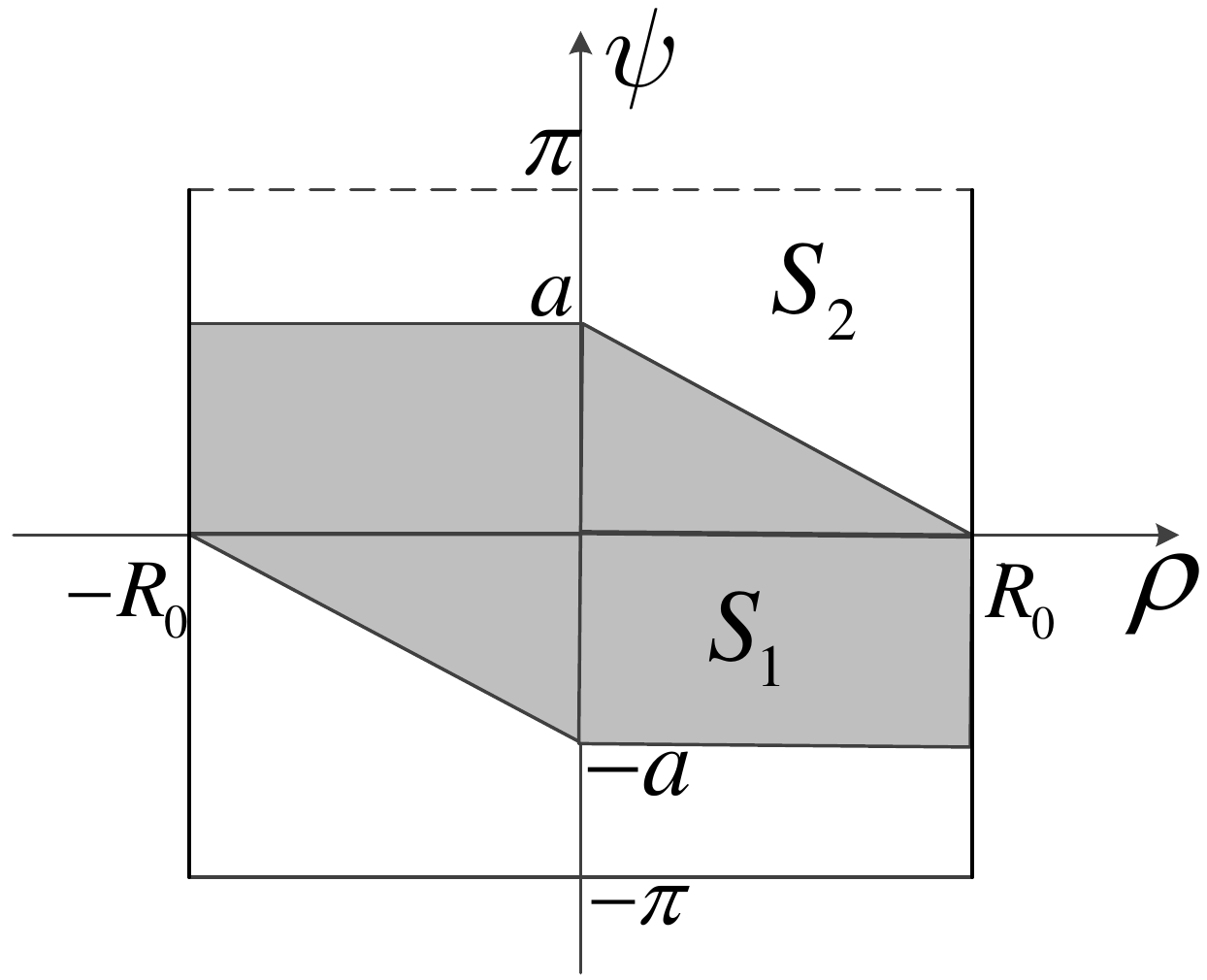}
		\caption{Set $\mathcal{S}$ with its two partitioned subsets $\mathcal{S}_1$ and $\mathcal{S}_2$ in~\cite{YingLan}.}
		\label{fig2}
	\end{figure}

	When ignoring the speed constraints, we can extract a set $\mathcal{S}$ as $\mathcal{S}=\{(\rho,\psi)\colon\rho\in[-R_0,R_0],\psi\in[-\pi,\pi)\}$, which is illustrated in Fig.~\ref{fig2}, as defined in~\cite{YingLan}, where $\mathcal{S}$ is partitioned into two parts $\mathcal{S}_1$ and $\mathcal{S}_2$.
	$\mathcal{S}_1$ is the coordination set, defined as $\mathcal{S}_1 := \{(\rho,\psi)\in\mathcal{S}\colon |\rho|\leq R_0, |\psi|\leq a, |a\rho+R_0\psi|\leq aR_0\}$, and $\mathcal{S}_2 := \mathcal{S}\setminus\mathcal{S}_1$.
	Parameter $a$ satisfies $0<a<\min\{(\pi/2),R_0\}$.
	It has been proved that, by applying an appropriate hybrid control law, the path following error $\phi_i$ of the $i$\textsuperscript{th} UAV, whose dynamics are represented by~\eqref{eq_patherror} without constraints represented by~\eqref{eq_constraint}, converges to $\mathbf{0}$.
	To be more specific, if $\phi_i$ is in $\mathcal{S}_2$ initially, then it cannot leave $\mathcal{S}$ and will enter $\mathcal{S}_1$ in a finite time, and finally converge to $\mathbf{0}$, with the arc distance $\zeta_i$ approaching $L$ as $t\rightarrow\infty$.
	In this process, when $\phi_i(t)\in\mathcal{S}_2$, the $i$\textsuperscript{th} UAV only works at single-agent level (i.e., without any coordination), and the coordination algorithm gets to work only after $\phi_i$ gets into the coordination set.
	Unfortunately, when considering the speed constraints represented by~\eqref{eq_constraint}, the above results cannot be guaranteed.
	
	\begin{remark}
		We can always characterize a small region $\mathcal{S}_e$, such that when $\phi_i(t_0)\in\mathcal{S}_e$, given any control input $(v_i,\omega_i)\in[v_{\min},v_{\max}]\times[-\omega_{\max},\omega_{\max}]$, there exists a time $t_1>t_0$ such that $\phi_i(t_1)\notin\mathcal{S}$.
		Without loss of generality, we assume the UAVs follow a path with the curvature $\kappa(p_i)\in(-\kappa_0,0]$.
		We define $\mathcal{S}_e=\{\phi_i\in\mathcal{S}\colon \frac{v_{\min}(\psi_i-\epsilon_0)\sin\epsilon_0 }{\omega_{\max}}+\rho_i-R_0 > 0, \rho_i\in[0,R_0], \psi_i\in[0,\pi/2]\}$, where $0<\epsilon_0<\pi/2$.
		We note that $\mathcal{S}_e$ is not an empty set, otherwise, $\frac{v_{\min}(\psi_i-\epsilon_0) \sin\epsilon_0 }{\omega_{\max}}+\rho_i-R_0\leq 0$ should hold for all $\rho_i\in[0,R_0]$ and $\psi_i\in[0,\pi/2]$, which can be verified to be impossible by taking $\rho_i=R_0$ and $\psi_i=\pi/2$.
		According to~\eqref{eq_patherror}, when $\psi_i\in[\epsilon_0,\pi/2]$, $\dot{\psi}_i=\omega_i-\frac{\kappa(p_i)v_i\cos\psi_i}{1-\kappa(p_i)\rho_i} \geq-\omega_{\max}$. Suppose $t^*$ is the minimum time it takes to satisfy $\psi_i\leq\epsilon_0$, then $t^*\geq \frac{\psi_i(t_0)-\epsilon_0}{\omega_{\max}}$. When $t\in[t_0,t_0+t^*]$, we have $\dot{\rho}_i=v_i\sin\psi_i\geq v_{\min}\sin\epsilon_0>0$, and thus $\dot{\rho}_i\cdot\frac{\psi_i(t_0)-\epsilon_0}{\omega_{\max}}\geq \frac{v_{\min}(\psi_i(t_0)-\epsilon_0)\sin\epsilon_0 }{\omega_{\max}} > R_0-\rho_i(t_0)$, meaning $\phi_i$ will leave $\mathcal{S}$ before $t_0+t^*$.
		Therefore, it is impossible to use the results in~\cite{YingLan} to solve the speed constrained coordinated path following problem defined in Problem~\ref{prob:Coordinated Path Following}, since not all the points in $\mathcal{S}$ can be guaranteed always within $\mathcal{S}$ even under any possible control input.
	\end{remark}

	To solve Problem~\ref{prob:Coordinated Path Following}, we have to further partition set $\mathcal{S}$ to specify the invariant subsets and design the control laws accordingly.
	%
	%
	%
	To be more specific, we define a new coordination set and propose the corresponding control law in Section~\ref{sec:S1};
	and for the points outside the coordination set, we design a single-agent level control law in Section~\ref{sec:S2}.
	
		Before we proceed, the following lemma is necessary throughout the paper.
			
		\begin{lemma}[Pages 61-62 of~\cite{Hassan}]
			\label{lm_nagumo}
			Consider a simple closed contour defined by $g(x)=0$, with $g(x)<0$ enclosed by the contour, where $g(x)$ is a continuously differentiable function. The vector field $f(x)$ at a point $x$ on the coutour points inward if the inner product of $f(x)$ and the gradient vector $\bigtriangledown g(x)$ is negative, i.e., $f(x)\cdot\bigtriangledown g(x)<0$;
			and the vector field points outward if $f(x)\cdot\bigtriangledown g(x)>0$;
			and it is tangent to the contour if $f(x)\cdot\bigtriangledown g(x)=0$. The trajectory can leave the set enclosed by the contour, only if the vector field points outward at some point on its boundary, i.e., $\exists~x$ such that $g(x)=0$ and $f(x)\cdot\bigtriangledown g(x)>0$.
		\end{lemma}

	\section{Coordinated Control Law In Coordination Set\label{sec:S1}}
	
	In this section, we discuss how to control the UAVs, whose path following errors are within a given set (called the coordination set $\mathcal{S}_1$), to move along the path $\Gamma$ in a coordination manner.

	We formulate the coordination set as $\mathcal{S}_1=\{(\rho,\psi):|\rho|\leq R_1, |\psi|\leq a, |a\rho+R_1\psi|\leq aR_1\}$, where $R_1< R_0$. Since $\kappa_0=\frac{1}{R_0}$, we have $\kappa_0R_1<1$.
	We note that this newly defined coordination set is relatively smaller compared to that without considering the speed constraints (see Fig.~\ref{fig2}).
	With speed constraints, parameters $a$ and $R_1$ should be properly selected.
	Thus, we first illustrate the parameter selection (Section~\ref{sec:parmasel}) before designing the control law (Section~\ref{sec:Control Law in S1}).

	\subsection{Parameter Selection of $\mathcal{S}_1$\label{sec:parmasel}}
	
	The selection of parameters in $\mathcal{S}_1$ follows two basic principles, which are illustrated one by one as follows.

	\subsubsection{The First Principle} We need to guarantee the existence of a proper control law making $\mathcal{S}_1$ an \emph{invariant set} (see \cite{Hassan}, page 127), i.e., if the initial path following error $\phi_i(t_0)\in\mathcal{S}_1$ at time $t_0$, then $\phi_i(t)\in\mathcal{S}_1$ for any $t>t_0$. In that way, for each $\phi_i(t_0)\notin\mathcal{S}_1$, we only need to design the control law to guarantee $\phi_i(t)$ entering $\mathcal{S}_1$. 

	\begin{remark}
		Denote the set $\partial\mathcal{S}_1$, as the intersection of $\mathcal{S}_1$ with the following set: $ \{(\rho,\psi)\colon|\psi|=a\}\cup\{(\rho,\psi)\colon|a\rho+R_1\psi|=aR_1\}$.
		It can be seen that $\partial\mathcal{S}_1$ is a subset of the boundary of $\mathcal{S}_1$.
		According to Lemma~\ref{lm_nagumo}, the first principle to guarantee $\phi_i(t)$ will not leave $\mathcal{S}_1$ is equivalent to guarantee $\phi_i(t)$ will never get across $|a\rho+R_1\psi|=aR_1$ in the first and the third quadrants, and $|\psi|=a$ in the second and fourth quadrants. Since $v_i>0$, it is certain that $\phi_i(t)$ will never leave $\mathcal{S}_1$ from $|\rho|=R_1$ in the second and the fourth quadrants, and that is why we do not include set $\{(\rho,\psi)\colon|\rho|=R_1\}$ in $\partial\mathcal{S}_1$. 
		\label{rmk_whichescape}
	\end{remark}
	
	Then, we have the following lemma:
	\begin{lemma}
		A sufficient condition under the first principle is that, for any $\phi_i\in\partial\mathcal{S}_1$, there exist $v_i$ and $\omega_i$ satisfying~\eqref{eq_constraint}, such that one of the four conditions [i.e., one of the four inequalities~\eqref{eq_first}-\eqref{eq_fourth}] holds:
		\begin{equation}
		\begin{cases}
		v_i(a\sin\psi_i-R_1\frac{\kappa(p_i)\cos\psi_i}{1-\kappa(p_i)\rho_i})+R_1\omega_i+R_1\alpha\leq0,\\
		\rho_i\geq 0,~~\psi_i\geq 0;\\
		\end{cases}
		\label{eq_first}
		\end{equation}
		
		\begin{equation}
		\begin{cases}
		\omega_i-\frac{\kappa(p_i)v_i\cos\psi_i}{1-\kappa(p_i)\rho_i}+\alpha\leq0,\\
		\rho_i< 0,~~\psi_i> 0;\\
		\end{cases}
		\label{eq_second}
		\end{equation}	
		
		\begin{equation}
		\begin{cases}
		v_i(a\sin\psi_i-R_1\frac{\kappa(p_i)\cos\psi_i}{1-\kappa(p_i)\rho_i})+R_1\omega_i-R_1\alpha\geq0,\\
		\rho_i\leq 0,~~\psi_i\leq 0;\\
		\end{cases}
		\label{eq_third}
		\end{equation}				
		
		\begin{equation}
		\begin{cases}
		\omega_i-\frac{\kappa(p_i)v_i\cos\psi_i}{1-\kappa(p_i)\rho_i}-\alpha\geq0,\\
		\rho_i> 0,~~\psi_i< 0;\\
		\end{cases}
		\label{eq_fourth}
		\end{equation}
		\label{lm_first}
	where $\alpha$ is a small positive number.
	\end{lemma}

	Lemma~\ref{lm_first} can be easily derived by using Lemma~\ref{lm_nagumo}.
	
	%
	
With Lemma~\ref{lm_first}, we can get an important inequality for the parameter selection, given in the following lemma.

	\begin{lemma}\label{lem:Sufficient Condition under the First Principle 2}
		If there exists $v_m\in(v_{\min},v_{\max}]$ making~\eqref{eq6} and~\eqref{eq_ineqprinciple2} hold, then for any $\phi_i\in\partial\mathcal{S}_1$, there exist $v_i$ and $\omega_i$ satisfying~\eqref{eq_constraint}, such that one of the four inequalities~\eqref{eq_first}-\eqref{eq_fourth} holds.
		\begin{equation}
		\label{eq6}
		\sqrt{(\frac{a}{R_1})^2+\kappa^2_0}+\frac{\alpha}{v_m}\leq\frac{\omega_{\max}}{v_m},
		\end{equation}
	 \begin{equation}
			\label{eq_ineqprinciple2}
			\frac{\kappa_0}{1-\kappa_0R_1}+\frac{\alpha}{v_m}\leq\frac{\omega_{\max}}{v_{m}}.		
	\end{equation}
					
	\end{lemma}
	\begin{IEEEproof}
		See Appendix~\ref{apx:Proof of Lemma lem:Sufficient Condition under the First Principle 2}.
	\end{IEEEproof}
	
	
	\subsubsection{The Second Principle} We need to guarantee the existence of a proper control law, which ensures the sequence of the UAVs along the path to be fixed, once all the UAVs enter $\mathcal{S}_1$.
	More specifically, if UAV $i$ and its pre-neighbor UAV $j$ satisfy $\phi_i(t_0), \phi_j(t_0)\in\mathcal{S}_1$ at time $t_0$, then there will be no any other $k$\textsuperscript{th} UAV ($k\neq i, k\neq j$) turning to be the pre-neighbor of the $i$\textsuperscript{th} UAV from then on.
	Mathematically, if $\phi_i(t_0)\in\mathcal{S}_1,~\forall~i=1,\ldots,n$, then $\zeta_i(t)>0,~\forall~t>t_0$.\footnote{ We note that when $\zeta_i(t)$ is reduced to zero, it means the $i$\textsuperscript{th} UAV is being overtaken by another UAV, or is overtaking another UAV at time $t$.}

	\begin{lemma}
		The second principle holds if
		\begin{equation}
		\label{eq7}
		\frac{1}{1-\kappa_0R_1}v_{\min}+c\leq \frac{\cos a}{1+\kappa_0R_1}v_m,
		\end{equation}
		where $c>0$.
		\label{lm_second}
	\end{lemma}
	\begin{IEEEproof}
		See Appendix~\ref{apx:Proof of Lemma lm_second}.
	\end{IEEEproof}
	
	Besides inequalities~\eqref{eq6}-\eqref{eq7} (deduced from Lemma~\ref{lem:Sufficient Condition under the First Principle 2} and Lemma~\ref{lm_second}, respectively), there are some additional constraints for parameters:
	\begin{equation}
	\label{eq8}
	\begin{cases}
	0<a<\pi/2,\\
	0<R_1< R_0,\\
	v_{\min}< v_m\leq v_{\max},
	\end{cases}
	\end{equation}
	where $v_{\min}$, $v_{\max}$ and $\omega_{\max}$ are determined by the dynamics of the UAVs;
	$\kappa_0$ is determined by the path for the UAVs to follow; $c$ is chosen by the designers. Then, there are three parameters in~\eqref{eq8} left unknown, namely, $a$, $R_1$ and $v_m$.
	The selection of $a$, $R_1$ and $v_m$ can be regarded as an optimization problem.
	Its objective is to make $\mathcal{S}_1$ as large as possible, since a larger $\mathcal{S}_1$ contains more path following errors, which means more situations can be directly\footnote{We note that if a path following error is outside $\mathcal{S}_1$, we have to make it enter $\mathcal{S}_1$, before the coordination control law in Section~\ref{sec:Control Law in S1} is applied.} dealt with our proposed coordination control law in Section~\ref{sec:Control Law in S1}.
	%
	%
	This optimization problem is described by
	\begin{eqnarray}\label{eqobj}
	&\mathrm{maximize} &aR_1\\
	& \mathrm{s.t.} & \nonumber  \text{inequalities~\eqref{eq6}-\eqref{eq8} hold.}
	\end{eqnarray}

	If constraints~\eqref{eq6}-\eqref{eq8} are satisfied (see Lemma~\ref{lem:A Sufficient Condition to Guarantee the Feasibility of eqobj}), we can guarantee the existence of proper control law satisfying the aforementioned two principles.
	By solving the optimization problem~\eqref{eqobj}, we can select the optimized parameters $a$ and $R_1$ to design $\mathcal{S}_1$, and we can also obtain $v_m$, which is an important parameter in the control law proposed in Section~\ref{sec:Control Law in S1}.

	\begin{lemma}\label{lem:A Sufficient Condition to Guarantee the Feasibility of eqobj}
		A sufficient condition to guarantee the existence of feasible solutions of~\eqref{eqobj} is
		\begin{equation}\label{eqparam}
		\begin{cases}
		\kappa_0\leq\frac{\omega_{\max}}{v_{\max}},\\
		v_{\min}+c\leq v_{\max}.
		\end{cases}
		\end{equation}
		\label{lb_c_1}
	\end{lemma}

	\begin{remark}\label{rek:Parameter - c}
		Lemma~\ref{lb_c_1} can be verified easily. From it, we can get one guideline to choose the user-determined value of parameter $c$ in~\eqref{eqobj} as $c\leq v_{\max}-v_{\min}$.
		
		It is obvious from~\eqref{eq7} that the smaller $c$ is, the larger the coordination set will become, since we will get a larger feasible region in~\eqref{eqobj}.
		However, in terms of the control law designs, $c$ is not the smaller the better, because $c$ is also correlated to the convergence rate (more details are given in Section~\ref{sec:Control Law in S1}), and a trade-off should be made between the coordination set's size and the convergence rate.
	\end{remark}
	
	Above completes the parameter selection process for coordination set $\mathcal{S}_1$.
	With the parameter selection, we cannot only guarantee the existence of the control law satisfying the proposed two principles, but also design a proper coordination set $\mathcal{S}_1$.
	Next, we will design the control law, and show the importance of the two principles to the convergence of the path following error and the sequenced inter-UAV arc distance.

	\subsection{Control Law in $\mathcal{S}_1$}\label{sec:Control Law in S1}
\begin{algorithm}[!htb]
	\caption{Coordinated Path Following Control Law in $\mathcal{S}_1$}		
	\label{alg}			
	\footnotesize
	\begin{algorithmic}[1]
		\Require $\rho_i,\psi_i,\kappa(p_i),\zeta_i$\label{ln_obtain}
		\Ensure $v_i,\omega_i$
		\Procedure {CoordControl}{$\rho_i,\psi_i,\kappa(p_i),\zeta_i$}
		\State Set the forward speed as $$v_i=\mathrm{Sat}\Big(\frac{1-\kappa(p_i)\rho_i}{\cos\psi_i}\chi(\zeta_i),v_{\min},v_{\max}\Big);$$\label{ln_vi1}
		\State Set the angular speed as
		$$\omega_i=\mathrm{Sat}(\omega_d,-\omega_{\max},\omega_{\max}),$$ where
		$$\omega_d=v_i\left[-\frac{k_1\vartheta_i}{k_2}+\frac{\kappa(p_i)\cos\psi_i}{1-\kappa(p_i)\rho_i}\right]-\alpha\cdot\mathrm{sign}(\vartheta_i),$$ and $k_1>0,~k_2,k_3\geq1$ and $a\leq R_1k_1<ak_2$; \label{ln_omegai}
		\State $v_i\gets$\Call{ReSetValue}{$v_i,\omega_i,\rho_i,\psi_i,\kappa(p_i),\zeta_i$}
		\State \Return{$v_i,\omega_i$} 			
		\EndProcedure
		\vspace{1em}
		\Procedure {ReSetValue}{$v_i,\omega_i,\rho_i,\psi_i,\kappa(p_i),\zeta_i$}
		\If {$\phi_i\in\mathcal{S}^1_1$ \textbf{and} inequality~\eqref{eq_first} does not hold}
		\State
		$v_i=-\left[a\sin\psi_i-R_1\frac{\kappa(p_i)\cos\psi_i}{1-\kappa(p_i)\rho_i}\right]^{-1}R_1(\omega_{i}+\alpha)$
		\EndIf
		\If{$\phi_i\in\mathcal{S}^2_1$ \textbf{and} inequality~\eqref{eq_second} does not hold}
		\State
		$
		v_i=\frac{1-\kappa(p_i)\rho_i}{\kappa(p_i)\cos\psi_i}(\omega_{i}+\alpha)$
		\EndIf
		\If {$\phi_i\in\mathcal{S}^3_1$ \textbf{and} inequality~\eqref{eq_third} does not hold}
		\State $v_i=-\left[a\sin\psi_i-R_1\frac{\kappa(p_i)\cos\psi_i}{1-\kappa(p_i)\rho_i}\right]^{-1}R_1(\omega_{i}-\alpha)$
		\EndIf
		\If{$\phi_i\in\mathcal{S}^4_1$ \textbf{and} inequality~\eqref{eq_fourth} does not hold}
		\State $			
		v_i=\frac{1-\kappa(p_i)\rho_i}{\kappa(p_i)\cos\psi_i}(\omega_{i}-\alpha)$			
		\EndIf	
		\If{$\phi_i\in\mathcal{S}^5_1$ \textbf{and} $\omega_i-\frac{\kappa(p_i)v_i\cos\psi_i}{1-\kappa(p_i)\rho_i}-\alpha<0$}
		\State $
		v_i=\frac{1-\kappa(p_i)\rho_i}{\kappa(p_i)\cos\psi_i}(\omega_{i}-\alpha)$
		\EndIf			
		\If{$\phi_i\in\mathcal{S}^6_1$ \textbf{and} $\omega_i-\frac{\kappa(p_i)v_i\cos\psi_i}{1-\kappa(p_i)\rho_i}+\alpha>0$}
		\State $
		v_i=\frac{1-\kappa(p_i)\rho_i}{\kappa(p_i)\cos\psi_i}(\omega_{i}+\alpha)
		$
		\EndIf					
		\State \Return{$v_i$}
		\EndProcedure
	\end{algorithmic}
\end{algorithm}

	Now, we propose the coordinated path following control law in $\mathcal{S}_1$, which is included in Algorithm~\ref{alg}.
	We assume $\rho_i$, $\psi_i$ and $\kappa(p_i)$ can be calculated by the UAV itself, since the path is globally available to the UAV. Besides, $\zeta_i$ can be measured by the UAV itself or through communication with its pre-neighbor, and thus the multi-UAV coordination can be achieved through sensing or communication. When the $i$\textsuperscript{th} UAV does not have a pre-neighbor, we artificially set $\zeta_i=L$, i.e., the coordination error $L-\zeta_i$ is set to be zero.
	
	Before illustrating our control algorithm, we introduce a continuous function $\chi(\zeta_i)$, which is used in Line~\ref{ln_vi1} with the following properties:
	\begin{itemize}
		\item[i)] $\chi(\zeta_i)=\frac{1}{1-\kappa_0 R_1}v_{\min}$, when $\zeta_i\in[0,L-\delta_1)$, where $0<\delta_1<L$;
		\item[ii)] $\chi(L)=\frac{\lambda}{1-\kappa_0 R_1}v_{\min}+\frac{(1-\lambda)\cos a}{1+\kappa_0R_1}v_m$, where $0<\lambda<1$;
		\item[iii)] $\chi(\zeta_i)$ is non-decreasing when $\zeta_i\in[0,\infty)$, and is strictly increasing in $[L-\delta_2,L+\delta_2]$, where $0<\delta_2\leq\delta_1$.
	\end{itemize}
	
	We also use a saturation function $\mathrm{Sat}(\cdot)$ in Lines~\ref{ln_vi1} and~\ref{ln_omegai}, which is defined as follows (suppose $a<b$): $\mathrm{Sat}(x,a,b)=a$ if $x\leq a$, $\mathrm{Sat}(x,a,b)=x$ if $a<x\leq b$, and $\mathrm{Sat}(x,a,b)=b$ if $x>b$.

	The main procedure of the coordinated path following algorithm in $\mathcal{S}_1$, \algoname{CoordControl}, can be described as follows: in Line~\ref{ln_vi1}, we set the value of $v_i\in[v_{\min},v_{\max}]$ based on the coordination error.
	We calculate $\omega_i\in[-\omega_{\max},\omega_{\max}]$ in Line~\ref{ln_omegai}, which is a function of $v_i$ derived in Line~\ref{ln_vi1}.
	After calculating $v_i$ and $\omega_i$, we check whether some properties (see procedure~\algoname{ReSetValue}) hold with the derived $v_i$ and $\omega_i$, and if not, we will recalculate $v_i$ again. The checking-recalculating process is depicted by procedure \algoname{ReSetValue}. In this procedure, we divide $\mathcal{S}_1$ into six subsets, and reset the value of $v_i$ by using different rules for different subsets.

\begin{figure}[!htb]
		\centering
		\includegraphics[height=0.8in]{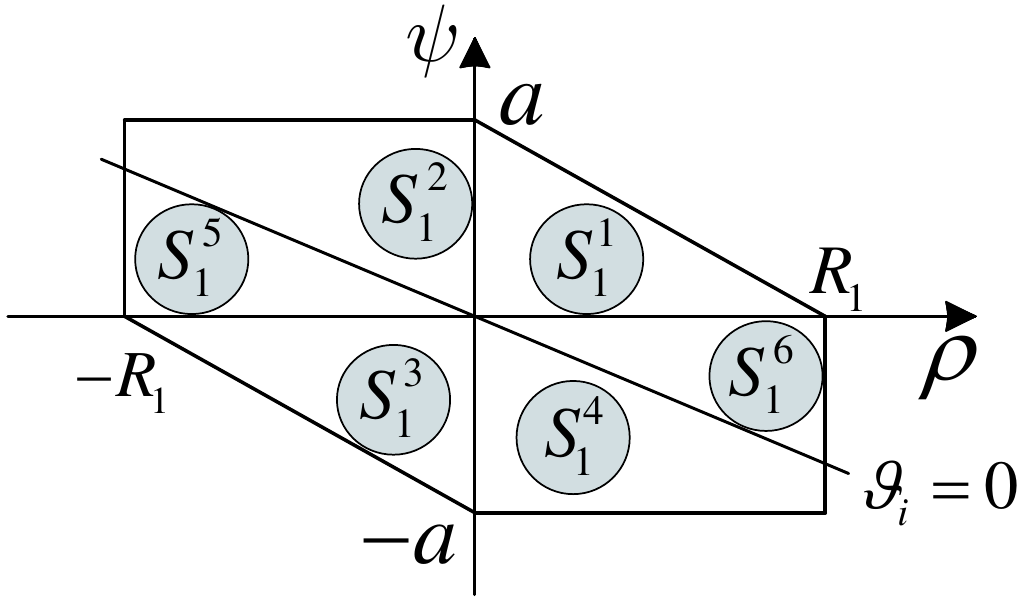}
		\caption{Six subsets of $\mathcal{S}_1$ divided by $\vartheta_i=0$, $\rho$-axis, and $\psi$-axis. 
		}
		\label{fig_VF}
	\end{figure}

	Let $\vartheta_i=k_1\rho_i+k_2\psi_i+k_3\sin\psi_i$, then $\vartheta_i=0$ is a nearly-straight curve passing through the origin, $\vartheta_i>0$ is on the upper-right side of the curve, and $\vartheta_i<0$ on its lower-left side. $\vartheta_i=0$ together with $\rho$-axis and $\psi$-axis divides $\mathcal{S}_1$ into six subsets as shown in Fig.~\ref{fig_VF}, which are defined as follows:
	$$
	\begin{aligned}	
	\mathcal{S}^1_1&=\{\phi_i\in\mathcal{S}_1\colon\rho_i>0,\psi_i\geq0,\vartheta_i>0\};\\
	\mathcal{S}^2_1&=\{\phi_i\in\mathcal{S}_1\colon\rho_i\leq0,\psi_i\geq0,\vartheta_i\geq0\};\\
	\mathcal{S}^3_1&=\{\phi_i\in\mathcal{S}_1\colon\rho_i<0,\psi_i\leq0,\vartheta_i<0\};\\
	\mathcal{S}^4_1&=\{\phi_i\in\mathcal{S}_1\colon\rho_i\geq0,\psi_i\leq0,\vartheta_i\leq0\};\\
	\mathcal{S}^5_1&=\{\phi_i\in\mathcal{S}_1\colon\rho_i<0,\psi_i>0,\vartheta_i<0\};\\
	\mathcal{S}^6_1&=\{\phi_i\in\mathcal{S}_1\colon\rho_i>0,\psi_i<0,\vartheta_i>0\}.\\
	\end{aligned}
	$$
	We note that according to the above partition, the origin is contained both in $\mathcal{S}^2_1$ and in $\mathcal{S}^4_1$. 

	In \algoname{ReSetValue}, we check whether one of the four inequalities~\eqref{eq_first}-\eqref{eq_fourth} holds when $\phi_i\in\mathcal{S}^1_1\cup\mathcal{S}^2_1\cup\mathcal{S}^3_1\cup\mathcal{S}^4_1$. If not, we will recalculate $v_i$.
	It is easy to find that, the recalculated $v_i$ can finally make one of the inequalities~\eqref{eq_first}-\eqref{eq_fourth} hold when $\phi_i\in\mathcal{S}^1_1\cup\mathcal{S}^2_1\cup\mathcal{S}^3_1\cup\mathcal{S}^4_1$. 
	We also note that there is no need to check inequalities~\eqref{eq_first}-\eqref{eq_fourth} when $\phi_i\in\mathcal{S}^5_1\cup\mathcal{S}^6_1$, since $\phi_i$ will not leave $\mathcal{S}_1$ from $|\rho_i|=R_1$ (see Remark~\ref{rmk_whichescape}). But we will make sure $\dot{\psi}_i\geq0$ when $\phi_i\in\mathcal{S}_1^5$, and $\dot{\psi}_i\leq0$ when $\phi_i\in\mathcal{S}_1^6$.

	There is an important property of procedure \algoname{ReSetValue}.
	\begin{lemma}
		\label{lm_vi}
		If $v_i$ is changed in \algoname{ReSetValue}, denote $v_{i1}$ as the forward speed calculated in Line~\ref{ln_vi1}, 
		and $v_i$ is the final returned value from \algoname{ReSetValue}, then $v_m\leq v_i< v_{i1}$.
		\begin{IEEEproof}
		See Appendix~\ref{apx:Proof of Lemma lm_vi}.
		\end{IEEEproof}
	\end{lemma}

According to the definition of function $\mathrm{Sat}(\cdot)$, we have $v_{i1}\leq v_{\max}$.
Therefore, Lemma~\ref{lm_vi} implies that, control inputs $v_i$ and $\omega_i$ calculated by Algorithm~\ref{alg} satisfy~\eqref{eq_constraint}.

We now show that the first principle is guaranteed by using Algorithm~\ref{alg}, which means $\mathcal{S}_1$ is made an invariant set.

	\begin{theorem}[Validation for the First Principle]
		%
		According to Algorithm~\ref{alg}, $v_i$ and $\omega_i$ make $\mathcal{S}_1$ an invariant set, i.e., if $\phi_i(t_0)\in\mathcal{S}_1$, then $\phi_i(t)\in\mathcal{S}_1$, $\forall~t>t_0$.		
		\label{thm_invariant}
	\end{theorem}
	
	\begin{IEEEproof}
		It is easy to check that procedure~\algoname{ReSetValue} can guarantee one of the four inequalities~\eqref{eq_first}-\eqref{eq_fourth} hold when $\phi_i\in\mathcal{S}^1_1\cup\mathcal{S}^2_1\cup\mathcal{S}^3_1\cup\mathcal{S}^4_1$.
			According to the definition of these sets,  $\partial\mathcal{S}_1\subsetneq\mathcal{S}^1_1\cup\mathcal{S}^2_1\cup\mathcal{S}^3_1\cup\mathcal{S}^4_1$. 
			%
			 With Lemma~\ref{lm_first}, we can conclude that $\mathcal{S}_1$ is an invariant set.
%
	\end{IEEEproof}

\begin{remark}\label{rek:Some Necessary Results}
We get the following results with Algorithm~\ref{alg}.
	\begin{itemize}
		\item[i)] If $\vartheta_i>0$, then $\dot{\psi}_i\leq-\alpha$; if $\vartheta_i<0$, then $\dot{\psi}_i\geq\alpha$. 
		\item[ii)] If $\phi_i\in\mathcal{S}^1_1\cup\mathcal{S}^3_1$, and $\psi_i\neq 0$, then $\frac{\dot\psi_i}{\dot\rho_i}\leq-\frac{a}{R_1}$.
	\end{itemize}	
\end{remark}
	
	
	In terms of the second principle proposed in Section~\ref{sec:parmasel}, Theorem~\ref{thmarcdist} illustrates that Algorithm~\ref{alg} guarantees the sequence of UAVs becoming fixed, once all the UAVs enter $\mathcal{S}_1$.
	
	\begin{theorem}[Validation for the Second Principle]\label{thmarcdist}
		%
		Suppose $\phi_i(t_0)\in\mathcal{S}_1,~\forall~i=1,\ldots,n$. By executing Algorithm~\ref{alg}, if $\zeta_i(t_0)>0$, then $\zeta_i(t)>0$ holds, $\forall~t\geq t_0$.
	\end{theorem}
	\begin{IEEEproof}
		Firstly, each UAV's speed along the path $v_i^r$ is not smaller than $\chi(0)$ with Algorithm~\ref{alg}.
		This is obvious if the value of $v_i$ is not changed in~\algoname{ReSetValue} by using the definition of $\chi(\cdot)$.
		If the value of $v_i$ is changed in \algoname{ReSetValue}, according to Lemma~\ref{lm_vi}, $v_i\geq v_{m}$, then $v^r_i\geq v_m\frac{\cos a}{1+\kappa_0R_1}> \chi(0)$.
		
		Now let the $j$\textsuperscript{th} UAV be the pre-neighbor of the $i$-th UAV, then $\dot{\zeta}_i=v^r_j-v^r_i$. Suppose initially $0<\zeta_i(t_0)\leq L-\delta_1$, and thus $v^r_i=\chi(0)$.
		Using similar analyses with Lemma~\ref{lm_second}, it can be concluded that
		%
 $\zeta_i(t)>0$, $\forall~t\geq t_0$.		
	\end{IEEEproof}
	
	Now we show the importance of the two principles to the convergence of the path following error and sequenced inter-UAV arc distance in $\mathcal{S}_1$. In terms of the path following error, we have the following theorem stating that all the UAVs will finally move on the desired path if they are all in $\mathcal{S}_1$ initially.
	\begin{theorem}[Convergence of Path Following Error]
		\label{thm_pathfollow}
		%
		 By executing Algorithm~\ref{alg}, if $\phi_i(t_0)\in\mathcal{S}_1$, then $\lim\limits_{t\rightarrow\infty}\phi_i(t)=\mathbf{0}$.
	\end{theorem}
	\begin{IEEEproof}
		Firstly, recall~i) in Remark~\ref{rek:Some Necessary Results}, $\dot{\psi}_i\leq-\alpha<0$ when $\vartheta_i>0$; and $\dot{\psi}_i\geq\alpha>0$ when $\vartheta_i<0$.
		We note that $\mathcal{S}_1$ is an invariant set according to Theorem~1, and $|\psi_i|\leq a$ when $\phi_i\in\mathcal{S}_1$.
		Therefore, for any $\phi_i(0)\in\mathcal{S}_1$, there exists a finite time $t_0\leq \frac{2a}{\alpha}$, such that $\vartheta_i(t_0)=0$.
		Secondly, by~i) in Remark~6, we can show that for any $\phi_i(t_0)\in\{(\rho_i,\psi_i)\colon\vartheta_i=0\}$, $\phi_i$ will not go to $\mathcal{S}^5_1$ or $\mathcal{S}^6_1$ directly, since $\vartheta_i\dot{\vartheta}_i=\vartheta_i(k_1\dot{\rho}_i+k_2\dot{\psi}_i+k_3\dot{\psi}_i\cos\psi_i)<0$ when $\phi_i\in\mathcal{S}_1^5\cup\mathcal{S}_1^6$.
		Thus if $\phi_i(t_0)\in \{(\rho_i,\psi_i)\colon\vartheta_i=0\}$, the possibilities for the movement of $\phi_i$ after $t_0$ can be partitioned into the following three cases.
		%
		
		\emph{Case 1:} $\phi_i(t)\in\{(\rho_i,\psi_i)\colon\vartheta_i=0\}$ holds for any $t\geq t_0$. 
		Since $\vartheta_i=k_1\rho_i+k_2\psi_i+k_3\sin\psi_i$, it can be rewritten as $\vartheta_i=k_1\rho_i+k_2\arcsin\frac{\dot{\rho}_i}{v_i}+k_3\frac{\dot{\rho}_i}{v_i}$. Denote $h(x)=k_2\arcsin\frac{x}{v_i}+k_3\frac{x}{v_i}, x\in[-v_i\sin a,v_i\sin a]$.
		Obviously, $h(x)$ is an odd function, and its inverse function $h^{-1}(x)$ exists.
		Thus if $\vartheta_i(t)\equiv0$ holds for all $t\geq t_0$, then $\dot{\rho}_i=-h^{-1}(k_1\rho_i)$.
		Since $h(x)$ is a Lipschitz function, there exists a positive constant $c_1$ such that $|h(x)|\leq |c_1x|$. Then $|\rho_i(t)|\leq |\rho_i(t_0)|\exp[-\frac{k_1}{c_1}(t-t_0)]$, meaning $\rho_i(t)\to 0$ as $t\to\infty$.
		Moreover, $\psi_i(t)\to 0$ since $\vartheta_i(t)=k_1\rho_i+k_2\psi_i+k_3\sin\psi_i\equiv0$ holds for all $t\geq t_0$.
		
		\emph{Case 2:} $\phi_i(t)\notin\{(\rho_i,\psi_i)\colon\vartheta_i=0\}$ for some $t>t_0$, but $\phi_i(t)\in\mathcal{S}^2_1\cup \mathcal{S}^4_1$, $\forall t>t_0$.
		In this case, since $\dot{\psi}_i\geq \alpha$ if $\phi_i\in \mathcal{S}^4_1 \setminus\{(\rho_i,\psi_i)\colon\vartheta_i=0\}$, and $\dot{\psi}_i\leq -\alpha$ if $\phi_i\in \mathcal{S}^2_1 \setminus\{(\rho_i,\psi_i)\colon\vartheta_i=0\}$. Therefore, $|\psi_i|$ is non-increasing when $\phi_i(t)\notin\{(\rho_i,\psi_i)\colon\vartheta_i=0\}$, and the total time duration that $\phi_i(t_0)\notin\{(\rho_i,\psi_i)\colon\vartheta_i=0\}$ is finite, which is no more than $\frac{a}{\alpha}$.
		Besides, $|\rho_i|$ is also non-increasing when $\phi_i(t)\notin\{(\rho_i,\psi_i)\colon\vartheta_i=0\}$, and
		by combining the convergence results in \emph{Case~1}, we get $|\rho_i(t)|\leq |\rho_i(t_0)|\exp[-\frac{k_1}{c_1}(t-t_0-\frac{a}{\alpha})]$, and thus $\lim\limits_{t\to\infty}\rho_i(t)=0$.
		Since $|\psi_i|$ is non-increasing and bounded, then $\lim\limits_{t\to\infty}|\psi_i|$ exists. 
		Note that the total time duration that $\phi_i(t_0)\notin\{(\rho_i,\psi_i)\colon\vartheta_i=0\}$ is finite and $\lim\limits_{t\to\infty}\rho_i(t)=0$, it leads to $\lim\limits_{t\to\infty}\psi_i(t)=0$, and thus $\lim\limits_{t\to\infty}\phi_i(t)=\mathbf{0}$.
		
		\begin{figure}[!htb]
			\centering
			\includegraphics[scale=0.42]{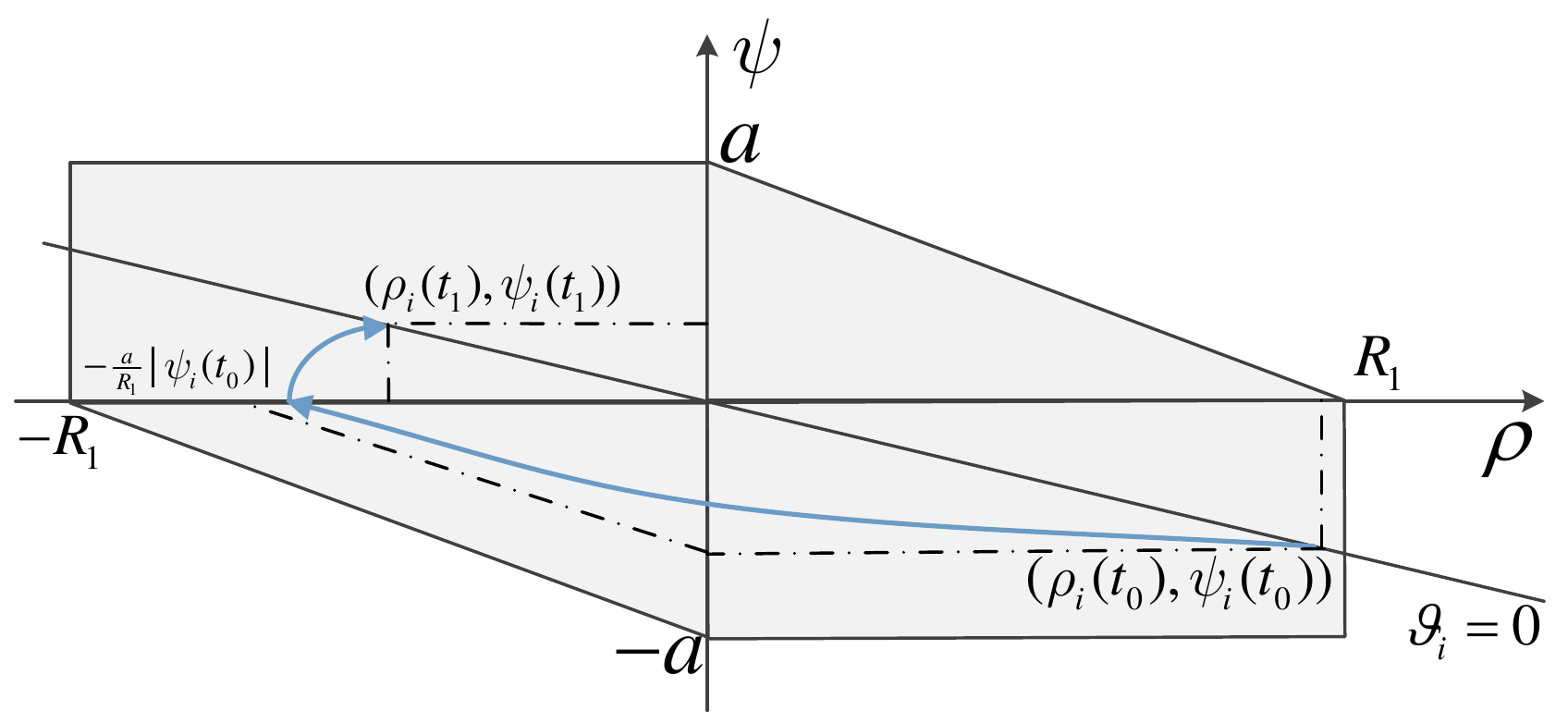}
			\caption{An illustration of the state trajectory in \emph{Case~3} when initially $\phi_i(t_0)\in\{(\rho_i,\psi_i)\colon\vartheta_i=0 \}$.}
			\label{fig_S1}
		\end{figure}
		
		\emph{Case 3:} there exists $\phi_i(t_0)\in\{(\rho_i,\psi_i)\colon\vartheta_i=0 \}$ and $\phi_i(t)\notin\mathcal{S}^2_1\cup \mathcal{S}^4_1$ for some $t>t_0$. Since $\phi_i$ will not go to $\mathcal{S}^5_1$ and $\mathcal{S}^6_1$ directly, then $\phi_i$ must have entered $\mathcal{S}^1_1\cup\mathcal{S}^3_1$ through the $\psi$-axis.
		Without loss of generality, we assume 
		$\rho_i(t_0)>0$ and $\psi_i(t_0)<0$, as shown in Fig.~\ref{fig_S1}.
		After $\phi_i$ enters $\mathcal{S}^3_1$, since $\dot{\psi}_i\geq\alpha>0$ holds in $\mathcal{S}^3_1$ and $\mathcal{S}^5_1$, there exists a finite time $t_1\leq t_0+\frac{2a}{\alpha}$ such that $\phi_i(t_1)\in\{(\rho_i,\psi_i)\colon\vartheta_i=0\}$, with $\rho_i(t_1)\leq0$ and $\psi_i(t_1)\geq0$.
		We use $(-r_1,0)$ to denote the intersection of the trajectory of $\phi_i$ with the $\rho$-axis, since $\dot{\psi}_i\geq \alpha>0$ when $\phi_i\in\mathcal{S}^4_1\setminus\{(\rho_i,\psi_i)\colon\vartheta_i=0\}$, and inequality~\eqref{eq_third} holds when $\phi_i\in\mathcal{S}^3_1$, which means $r_1<\frac{R_1}{a}|\psi_i(t_0)|$.
		Moreover, since $\dot{\rho}_i=v_i\sin\psi_i>0$ in $\mathcal{S}^5_1$. Therefore, $|\rho_i(t_1)|<r_1<\frac{R_1}{a}|\psi_i(t_0)|$.
		We note that $|\psi_i(t_0)|\leq \frac{k_1}{k_2}|\rho_i(t_0)|$ and $|\psi_i(t_1)|\leq \frac{k_1}{k_2}|\rho_i(t_1)|$. Thus $|\rho_i(t_1)| < \frac{R_1k_1}{ak_2}|\rho_i(t_0)|$ and $|\psi_i(t_1)| < \frac{R_1k_1}{ak_2}|\psi_i(t_0)|$. Let $\sigma:=\frac{R_1k_1}{ak_2}$, and we have $\sigma<1$.
		Therefore, $|\rho_i(t)|\leq |\rho_i(t_0)|$, $|\psi_i(t)|\leq |\psi_i(t_0)|$ when $t\in[t_0,t_1]$.
		By symmetry, when $\phi_i(t_1)\in\{(\rho_i,\psi_i)\colon\vartheta_i=0\}$, there are three possibilities for the movement of $\phi_i$ after $t_1$, corresponding to the three cases we listed here.
		We only consider \emph{Case~3}, since the first two cases implies the path following error converges to zero by our analysis above.
		Now if $\phi_i$ enters $\mathcal{S}_1^1$ after $t_1$ and then reaches $\{(\rho_i,\psi_i)\colon\vartheta_i=0\}$ again at time $t_2$, it follows that $|\rho_i(t)|\leq \sigma|\rho_i(t_0)|$, $|\psi_i(t)|\leq \sigma|\psi_i(t_0)|$ when $t\in[t_1,t_2]$.
		Proceeding forward, we get $|\rho_i(t)| \leq \sigma^{m-1} |\rho_i(t_1)|$, and $|\psi_i(t)| \leq \sigma^{m-1} |\psi_i(t_1)|,$ for $t\in[t_m,t_{m+1}]$, where $t_m$ corresponds to the $m$-th time that $\phi_i$ leaves $\mathcal{S}^2_1\cup\mathcal{S}^4_1$ and reaches $\{(\rho_i,\psi_i)\colon\vartheta_i=0\}$ again.
		Since $\sigma^{m-1}\to 0$ as $m\to \infty$, we get $\lim\limits_{t\rightarrow\infty} |\rho_i(t)| = 0$ and $\lim\limits_{t\rightarrow\infty} |\psi_i(t)| = 0$, i.e., $\lim\limits_{t\rightarrow\infty} \phi_i(t) = \mathbf{0}$.
	
		Combining the above cases together, we conclude that $\lim\limits_{t\rightarrow\infty} \phi_i(t) = \mathbf{0}$ always holds.
	\end{IEEEproof}
	
	\begin{remark}
		According to the proof of Theorem~\ref{thm_pathfollow}, if $\phi_i(t_0)\in \{(\rho_i,\psi_i)\colon\vartheta_i=0\}$, there are three possibilities for the movement of $\phi_i$ after $t_0$. However, in a small neighborhood of the origin, it satisfies $k_1v_{\max}|\sin\psi_i|\leq k_2\alpha+k_3\alpha\cos\psi_i$. As a result, $\vartheta_i\dot{\vartheta}_i\leq 0$ always holds in this neighborhood, and $\vartheta_i\dot{\vartheta}_i= 0$ if and only if $\vartheta_i=0$, meaning $\phi_i$ can only slide on $\vartheta_i=0$ once it reaches $\{(\rho_i,\psi_i)\colon\vartheta_i=0\}$ in this neighborhood.
		We also note that since the signum function is adopted in the control law, $\omega_i$ is not continuous at $\vartheta_i=0$, but $\phi_i$ is continuous. Thus, the solution should be understood in the sense of Filippov.
		To eliminate chattering caused by the signum function, a high-slope saturation function can be used to replace it~\cite{Hassan}.
	\end{remark}
	
	\begin{remark}
		It should be noted that the Lyapunov method can be employed in each subset of $\mathcal{S}_1$, but we still need to prove the overall stability by considering the six subsets and their boundaries altogether, which is actually what we did in our current proof of Theorem~\ref{thm_pathfollow}. To be more specific, consider the following Lyapunov function:
		\begin{equation*}
		V=\begin{cases}
		\begin{aligned}
		&\frac{1}{2}(k_1\rho_i+k_2\psi_i+k_3\sin\psi_i)^2,~&\phi_i\in\mathcal{S}^1_1\cup\mathcal{S}^3_1;\\
		&\frac{1}{2}(k_2\psi_i+k_3\sin\psi_i)^2,~&\phi_i\in\mathcal{S}^2_1\cup\mathcal{S}^4_1;\\
		&\frac{1}{2}(k_1\rho_i)^2,~&\phi_i\in\mathcal{S}^5_1\cup\mathcal{S}^6_1.\\
		\end{aligned}
		\end{cases}
		\end{equation*}
		It can be found that $V$ is a continuous function of $\phi_i$, and $\dot{V}<0$ in the interior of each subset. However, $V$ is not differentiable with respect to $\phi_i$ on the boundary of these subsets. Thus it calls for the similar technique to the above proof to analyze the boundaries in order to conclude the asymptotic stability.
	\end{remark}
	
	Now we have demonstrated the path following stability in $\mathcal{S}_1$. In terms of the coordination, we have the following claim that the coordination error will also converge to zero.

	\begin{theorem}[Convergence of Coordination Error]
		\label{thm_coordination}
		%
		 Suppose $\phi_i(t_0)\in\mathcal{S}_1$, $\forall~i=1,\ldots,n$, by executing Algorithm~\ref{alg}, $\lim\limits_{t\rightarrow\infty}\zeta_i(t)=L$ for all $i$.
	\end{theorem}
	\begin{IEEEproof}
		It follows from Theorem \ref{thmarcdist} that the pre-neighbor of each UAV does not change. Without loss of generality, suppose UAV 1 is the pre-neighbor of UAV 2, and UAV 2 is the pre-neighbor of UAV 3, and so on.
		%
		%
		Since UAV 1 does not have a pre-neighbor, it is artificially set as $\zeta_1=L$.
		For UAV 2, consider $V_2=(1/2)(\zeta_2-L)^2$, 
		thus $\dot{V_2}=(\zeta_2-L)\dot{\zeta_2}=(\zeta_2-L)(v^r_1-v^r_2)=-(\zeta_2-L)(v^r_2-\chi(L))$.
		According to the definition of $\chi(\cdot)$ and Lemma~\ref{lm_vi}, 
		 $(v^r_i-\chi(L))\cdot(\zeta_i-L)\geq0$, and the equality holds if and only if $\zeta_i=L$.
		Thus, we obtain $\lim\limits_{t\rightarrow\infty}\zeta_2(t)=L$ by using LaSalle's invariance principle~\cite{Hassan}.
		As a result, $\lim\limits_{t\to\infty}v_2^r=\chi(L)$.
		%
		%
		%
		Since $\dot{\zeta}_3=(v_2^r-\chi(L))-(v_3^r-\chi(L))$, and the system described by equation $\dot{\zeta}_i=-(v_i^r-\chi(L))$ will converge to $\lim\limits_{t\rightarrow\infty}\zeta_i=L$ (we get this claim by using the same analyses for UAV~2), with the Limiting Equation Theorem~\cite{Barkana2014}, we get $\lim\limits_{t\rightarrow\infty}\zeta_3=L$.
		Proceeding forward, we get $\lim\limits_{t\rightarrow\infty}\zeta_1(t)=\ldots=\lim\limits_{t\rightarrow\infty}\zeta_n(t)=L$.		
	\end{IEEEproof}
	\begin{remark}
		We have completed the control law design in the coordination set $\mathcal{S}_1$. We demonstrate that the designed control law can drive the UAVs whose path following errors initially in the coordination set to move onto the predefined path, with the inter-UAV arc distances converging to the desired value.
	\end{remark}
		
	Recall that in Remark~\ref{rek:Parameter - c}, we have proposed a guideline for the user-determined value $c$, that is, the smaller $c$ is, the larger the coordination set would become, since we will get a larger feasible region for~\eqref{eqobj}.
	However, the value of $c$ is not the smaller, the better.
	Now, we briefly analyze it from the perspective of convergence rate of the coordination error.
	Assume the $j$\textsuperscript{th} UAV is the pre-neighbor of the $i$\textsuperscript{th} UAV, and the coordination error of the $j$\textsuperscript{th} UAV is already zero, i.e., $\zeta_j=L$. If $\zeta_i<L$, then
	\begin{equation}\label{eqn:Pre-neighbor Inequality}
	\begin{aligned}
	|\dot{\zeta_i}|&=|v^r_i-v^r_j|=|\chi(\zeta_i)-\chi(\zeta_j)|\\
	&\leq(1-\lambda)[\frac{\cos a}{1+\kappa_0R_1}v_m-\frac{1}{1-\kappa_0 R_1}v_{\min}].\\
	\end{aligned}
	\end{equation}
	
	 Recall that~\eqref{eq7} is a precondition for~\eqref{eqobj}, then with a greater $c$, we are more likely to have a greater value of $\frac{\cos a}{1+\kappa_0R_1}v_m-\frac{1}{1-\kappa_0 R_1}v_{\min}$ from~\eqref{eqobj}, then we can get a greater upper-bound for the convergence rate of the coordination error in this case.
	Therefore, the determination of value $c$ is a trade-off between enlarging the coordination set and increasing the convergence rate of the sequenced inter-UAV arc distance.

	
	\section{Single-Agent Level Control Law Outside Coordination Set\label{sec:S2}}
	
	In Section~\ref{sec:S1}, we have designed the control law for those UAVs whose path following errors are within $\mathcal{S}_1$, such that they follow the path in a coordination manner.
	Now we discuss how to control the UAVs whose path following errors are outside $\mathcal{S}_1$.
	%
	%
	In that case, those UAVs do not follow the path cooperatively, but only adjust their path following errors at the single-agent level, in order to enter $\mathcal{S}_1$.



\begin{figure}[!htb]
		\centering
		\includegraphics[scale=0.3]{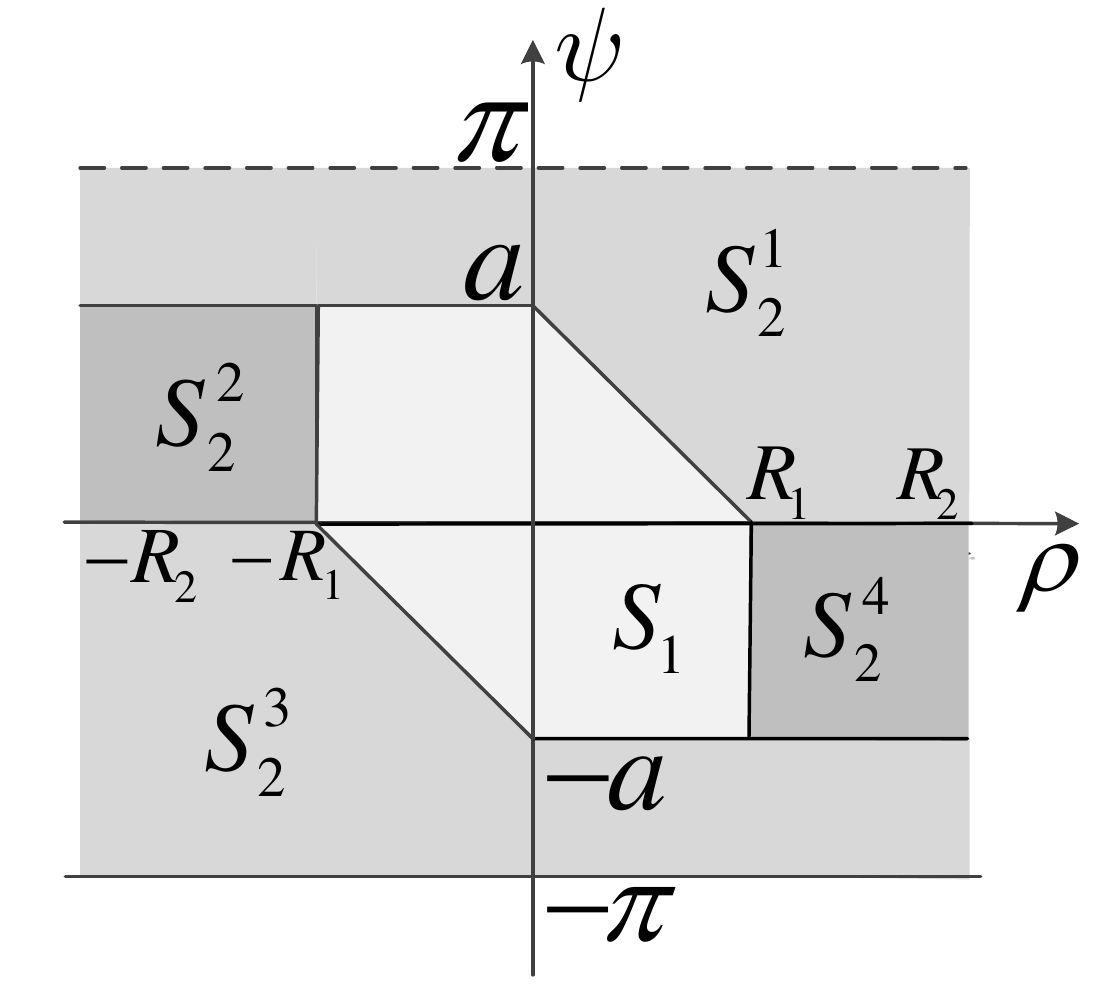}
		\caption{$\mathcal{S}_1$ and partition of $\mathcal{S}_2$.}
		\label{fig3}
	\end{figure}

	We label the set containing those path following errors outside $\mathcal{S}_1$ as $\mathcal{S}_2$.
	%
	To describe $\mathcal{S}_2$, we define a universe $\mathcal{S}$ which gives the scope for our designed control law\footnote{For the path following errors outside $\mathcal{S}$, they are distant away from the path $\Gamma$, and we can design Dubins paths to drive the errors into $\mathcal{S}$.}:
	\begin{equation}\label{eqn:Set S}
	\mathcal{S} = \left\{(\rho,\psi)\colon \rho\in[-R_2,R_2],\psi\in[-\pi,\pi)\right\}.
	\end{equation}
	Then, set $\mathcal{S}_2$ is defined as $\mathcal{S}_2 := \mathcal{S} \setminus \mathcal{S}_1$.
	We further divide $\mathcal{S}_2$ into four subsets: $\mathcal{S}^1_2$, $\mathcal{S}^2_2$, $\mathcal{S}^3_2$ and $\mathcal{S}^4_2$, as shown in Fig.~\ref{fig3}.
	%
	The mathematical descriptions of 
	these subsets are as follows:
	$$
	\begin{aligned}
	\mathcal{S}^1_2=&\Big(\{(\rho,\psi)\colon\psi>0\}\cap\mathcal{S}_2\setminus\mathcal{S}^2_2\Big)\bigcup\\
	&\{(\rho,\psi)\colon R_1<\rho\leq R_2, \psi=0\}, \\
	\mathcal{S}^2_2=&\{(\rho,\psi)\colon -R_2\leq\rho<-R_1,0<\psi\leq a\},\\
	\mathcal{S}^3_2=&\Big(\{(\rho,\psi)\colon\psi<0\}\cap\mathcal{S}_2\setminus\mathcal{S}^4_2\Big)\bigcup\\
	&\{(\rho,\psi)\colon-R_2\leq\rho<-R_1, \psi=0\},\\
	\mathcal{S}^4_2=&\{(\rho,\psi)\colon R_1<\rho\leq R_2,-a\leq\psi<0\}.
	\end{aligned}
	$$
	
	%
	%
	%
	%
	%

	In the rest of this section, we design the control laws for these four subsets:
	In Section~\ref{sec:Near Time Optimal Control Law}, we prove the existence of the control law which makes any path following error in sets $\mathcal{S}^2_2$ and $\mathcal{S}^4_2$ enter $\mathcal{S}_1$, and a near time optimal control law for sets $\mathcal{S}^2_2$ and $\mathcal{S}^4_2$ is designed.
	In Section~\ref{sec:Robust Control Law}, we design a robust control law for sets $\mathcal{S}^1_2$ and $\mathcal{S}^3_2$.


\subsection{Near Time Optimal Control Law in $\mathcal{S}^2_2$ and $\mathcal{S}^4_2$}\label{sec:Near Time Optimal Control Law}
	
	%
	Firstly, we prove the existence of control laws which makes the path following error within $\mathcal{S}^2_2\cup\mathcal{S}^4_2$ enter $\mathcal{S}_1$. 

	%
	%

	\begin{theorem}[Dynamics in $\mathcal{S}^2_2$ and $\mathcal{S}^4_2$]
		\label{thm_S2224}
		%
		If $ R_2<\frac{1}{\kappa_0}-\frac{v_{\min}}{\omega_{\max}}$, then for any $\phi_i(t_0)\in\mathcal{S}^2_2\cup\mathcal{S}^4_2$, there exists time $t_1$, and control $(v_i(t),\omega_i(t)),~t\in[t_0,t_1]$, satisfying constraint~(\ref{eq_constraint}), such that $\phi_i(t_1)\in\mathcal{S}_1$.
	\end{theorem}

	\begin{IEEEproof}
		Without loss of generality, we consider the condition in $\mathcal{S}^4_2$. The condition in $\mathcal{S}^2_2$ can be deduced similarly.
		
		In $\mathcal{S}^4_2$, since $\psi_i<0$, then $\dot\rho_i<0$.
		According to Lemma~\ref{lm_nagumo}, $\forall~\phi_i(t_0)\in\mathcal{S}^4_2$ would not leave $\mathcal{S}$ through $\rho=R_2$, 
		thus it can only enter $\mathcal{S}_1$ through $\rho=R_1$, or enter $\mathcal{S}^3_2$ through $\psi=-a$, or enter $\mathcal{S}^1_2$ through $\psi=0$, or remain in $\mathcal{S}^4_2$ forever.
		We show that by applying appropriate control law, the last three cases can be made impossible.
		Otherwise, suppose $\phi_i$ will enter $\mathcal{S}^3_2$, then there exists $\phi_i=(\rho_i,\psi_i)$ such that
		\begin{equation}
		\label{eq_contratry}
		\dot{\psi}_i=\omega_i-\frac{\kappa(p_i)v_i\cos\psi_i}{1-\kappa(p_i)\rho_i}<0
		\end{equation}
		holds for all $v_i$ and $\omega_i$ satisfying~\eqref{eq_constraint}.
		Let $\omega_i=\omega_{\max}$, $v_i=v_{\min}$, and~\eqref{eq_contratry} becomes:
		\begin{equation}
		\omega_{\max}-\frac{\kappa(p_i)v_{\min}\cos\psi_i}{1-\kappa(p_i)\rho_i}<0.
		\label{eq5}
		\end{equation}
		If $\phi_i\in\mathcal{S}^4_2$, then $\cos\psi_i>0$ and $1+\kappa(p_i)\rho_i>0$ hold, then 
		\begin{multline}
		\omega_{\max}-\frac{\kappa(p_i)v_{\min}\cos\psi_i}{1-\kappa(p_i)\rho_i}
		\geq \omega_{\max}-\frac{\kappa_0v_{\min}\cos\psi_i}{1-\kappa_0\rho_i}\\
		\geq \omega_{\max}-\frac{\kappa_0v_{\min}}{1-\kappa_0R_2}
		\geq0
		\end{multline}
		which is contrary to~\eqref{eq5}, meaning there exist $v_i$ and $\omega_i$ satisfying~\eqref{eq_constraint}, such that $\phi_i$ will not enter $\mathcal{S}^3_2$.
		In the same way, we conclude that there exist $v_i$ and $\omega_i$ such that $\phi_i$ will not enter $\mathcal{S}^1_2$ if $(\rho_i(t_0),\psi_i(t_0)\in\mathcal{S}_2^4$, and additionally, it can be made that $\psi_i(t)\leq\psi_i(t_0)$ for all $t\geq t_0$ before $\phi_i$ leaves $\mathcal{S}^4_2$. Thus 
		there exists 
		$t_1\leq \frac{R_1-\rho_i(t_0)}{v_{\min}\sin\psi_i(t_0)}$ such that $\phi_i(t_1)\in\mathcal{S}_1$.
	\end{IEEEproof}

	Since all the state errors in $\mathcal{S}^2_2$ and $\mathcal{S}^4_2$ can enter $\mathcal{S}_1$ when $ R_2<\frac{1}{\kappa_0}-\frac{v_{\min}}{\omega_{\max}}$, 
	 it is necessary to make $\phi_i$ enter $\mathcal{S}_1$ as soon as possible.
	Suppose $t_f$ is the minimum time instant such that $\phi_i(t_f)\in\mathcal{S}_1$, then the time optimal control objective is to minimize $t_f$, which can be formulated as follows. 
	$$
	\begin{aligned}
	\mathrm{(P1)}~~\mathrm{minimize} ~&J_1=t_f,\\
	\mathrm{s.t.}~&\phi_i(t_0)\in\mathcal{S}^2_2\cup\mathcal{S}^4_2, \phi_i(t_f)\in\mathcal{S}_1,\\
	& \text{and inequality~\eqref{eq_constraint} holds.}\\
	\end{aligned}
	$$
	
	In general, it is difficult to derive the optimal solution to P1. Here we adopt a greedy strategy to transform P1 into a near optimal control problem.
	Let $d=\inf_{\phi_f\in\mathcal{S}_1}\|\phi_i(t)-\phi_f\|$ denote the point-to-set Euclidean distance from $\phi_i(t)$ to $\mathcal{S}_1$.
	In our greedy strategy, the control objective is to minimize $\dot{d}$ at every time instant\footnote{This is why we say this strategy is ``greedy''.}, i.e., to drive $\phi_i(t)$ as close as possible towards $\mathcal{S}_1$. Thus P1 is transformed to the following problem.
	$$
	\begin{aligned}
	\mathrm{(P2)}~~\mathrm{minimize}~&J_2=\dot{d},\\
	&\text{where}~d=\inf_{\phi_f\in\mathcal{S}_1}\|\phi_i(t)-\phi_f\|,\\
	\mathrm{s.t.}~\phi_i(t)&\in\mathcal{S}^2_2\cup\mathcal{S}^4_2, ~\text{and inequality~\eqref{eq_constraint} holds.}\\
	\end{aligned}
	$$
	
	%
	
	%
	
	%

	%

	When $\phi_i(t)\in\mathcal{S}^4_2$, 
	$d=\rho_i-R_1$, then
	$$
	J_2=\dot{\rho_i}=v_i\sin\psi_i.
	$$
	
	We can see that $J_2$ is affine with respect to $v_i$. In $\mathcal{S}^4_2$, $\psi_i<0$, so we need to set $v_i=v_{\max}$. In terms of $\omega_i$, we have
	\begin{equation}
	\frac{\partial J_2}{\partial\omega_i}=\frac{\partial J_2}{\partial\psi_i}\cdot\frac{\partial\psi_i}{\partial\omega_i}
	=\frac{\partial J_2}{\partial\psi_i}\cdot\frac{\partial\dot\psi_i}{\partial\omega_i}\cdot dt
	=v_i\cos\psi_i\cdot dt>0.
	\end{equation}

	Therefore, to minimize $J_2$, we choose $\omega_i=-\omega_{\max}$, and the control law in $\mathcal{S}^4_2$ becomes
	\begin{equation}
	v_i=v_{\max},\quad
	\omega_i=-\omega_{\max}.
	\label{eq_law24norm}
	\end{equation}
	
	 With control law~\eqref{eq_law24norm}, we have
		\begin{multline}
		\dot{\psi}_i=\omega_i-\frac{\kappa(p_i)v_i\cos\psi_i}{1-\kappa(p_i)\rho_i}
		=-\omega_{\max}-\frac{\kappa(p_i)v_{\min}\cos\psi_i}{1-\kappa(p_i)\rho_i}\\
		\leq-\omega_{\max}+\frac{\kappa_0v_{\min}}{1+\kappa_0R_1}\leq0.
		\end{multline}

	According to Lemma~\ref{lm_nagumo}, by applying this control law, $\phi_i$ will not enter $\mathcal{S}^1_2$ through the $\rho$-axis.
	However, it may lead $\phi_i$ to enter $\mathcal{S}^3_2$.
 To prevent it, we set up a threshold $\epsilon_0$, where $0<\epsilon_0\ll a$. When $-a\leq\psi_i<-a+\epsilon_0$, we switch to a new control mode. There are two principles for this new mode. Firstly, we need $\dot\psi_i\geq0$ such that $\phi_i$ will not enter $\mathcal{S}^3_2$. Secondly, we need to minimize $\dot{d}$, i.e., $v_i\sin\psi_i$. Thus, the optimization problem P2 becomes~\eqref{eq_law24abnorm} when $-a\leq\psi_i<-a+\epsilon_0$:
	\begin{equation}
	\begin{aligned}
	\mathrm{minimize}~& v_i\sin\psi_i,\\
	\mathrm{s.t.}~~\omega_i-&\frac{\kappa(p_i)v_i\cos\psi_i}{1-\kappa(p_i)\rho_i}\geq 0,~ \text{and~\eqref{eq_constraint} holds.}\\
	\end{aligned}
	\label{eq_law24abnorm}
	\end{equation}
	
	Then, our near time optimal control law in $\mathcal{S}^4_2$ can be described as follows:
	\begin{enumerate}
		\item[i)]   If $\psi_i\geq-a+\epsilon_0$, the control law is~\eqref{eq_law24norm}.
		\item[ii)]   If $-a\leq\psi_i<-a+\epsilon_0$, the control law is the solution of~\eqref{eq_law24abnorm}, i.e.,
		\begin{itemize}
			\item   if $\omega_{\max}-\frac{\kappa(p_i)v_{\max}\cos\psi_i}{1-\kappa(p_i)\rho_i}\geq0$, then
			\begin{equation*}
			v_i=v_{\max},~
			\omega_i=\max\left\{-\omega_{\max},\frac{\kappa(p_i)v_{\max}\cos\psi_i}{1-\kappa(p_i)\rho_i}\right\};
			\end{equation*}
			\item   if $\omega_{\max}-\frac{\kappa(p_i)v_{\max}\cos\psi_i}{1-\kappa(p_i)\rho_i}<0$, then
			\begin{equation*}
			v_i=\frac{\omega_{\max}(1-\kappa(p_i)\rho_i)}{\kappa(p_i)\cos\psi_i},\quad
			\omega_i=\omega_{\max}.
			\end{equation*}
		\end{itemize}
	\end{enumerate}

	The control law in $\mathcal{S}^2_2$ can be designed in the same way as follows:
	\begin{enumerate}
		\item[i)]   If $\psi_i\leq a-\epsilon_0$, the control law is~\eqref{eq_law22norm};
		\begin{equation}
		v_i=v_{\max},\quad
		\omega_i=\omega_{\max}.
		\label{eq_law22norm}
		\end{equation}
		\item[ii)]   If $a-\epsilon_0<\psi_i\leq a$, the control law is the solution of~\eqref{eq_law22abnorm},
		\begin{equation}
		\begin{aligned}
		\mathrm{maximize}~& v_i\sin\psi_i,\\
		\mathrm{s.t.}~~\omega_i-&\frac{\kappa(p_i)v_i\cos\psi_i}{1-\kappa(p_i)\rho_i}\leq 0,~\text{and~\eqref{eq_constraint} holds.}\\
		\end{aligned}
		\label{eq_law22abnorm}
		\end{equation}
	\end{enumerate}
	It is easy to verify that the proposed near time optimal control law is the desired control in Theorem~\ref{thm_S2224}, i.e., $\phi_i$ will enter $\mathcal{S}_1$ when initially in $\mathcal{S}_2^2\cup\mathcal{S}_2^4$ by executing the control law.

	\subsection{Robust Control Law in $\mathcal{S}^1_2$ and $\mathcal{S}^3_2$}\label{sec:Robust Control Law}
	
	%
	By symmetric property, we only analyze the control law design in $\mathcal{S}^1_2$, and the methods can be applied to $\mathcal{S}^3_2$.

	It is obvious by Lemma \ref{lm_nagumo} that $\mathcal{S}^1_2$ is not an invariant set. 
	Specifically, there are five situations that $\phi_i(t)$ leaves $\mathcal{S}^1_2$:
	\begin{itemize}
		\item[i)] $\phi_i(t)$ passes through $\psi=a, -R_1\leq\rho<0$, or $a\rho+R_1\psi=aR_1, 0\leq\rho\leq R_1$, and enters $\mathcal{S}_1$ directly;
		\item[ii)] $\phi_i(t)$ passes through $\psi=a, -R_2\leq\rho<-R_1$, and enters $\mathcal{S}^2_2$;
		\item[iii)] $\phi_i(t)$ passes through $\psi=0, R_1<\rho\leq R_2$, and enters $\mathcal{S}^4_2$;
        \item[iv)] $\phi_i(t)$ passes through $\psi=\pi$, and enters $\mathcal{S}^3_2$;
		\item[v)] $\phi_i(t)$ passes through $\rho=R_2, 0\leq\psi\leq\pi$, and leaves $\mathcal{S}$.
		
	\end{itemize}
		
		Among all these situations, situation~i) is the case that $\phi_i(t)$ enters $\mathcal{S}_1$ directly.
		For situations~ii) and~iii), $\phi_i(t)$ enters $\mathcal{S}^2_2$ or $\mathcal{S}^4_2$, which can finally enter $\mathcal{S}_1$ (see Theorem~\ref{thm_S2224}).
		For situation~v), $\phi_i(t)$ leaves $\mathcal{S}$.
		For situation~iv), $\phi_i(t)$ enters $\mathcal{S}^3_2$, which has the same property as that in $\mathcal{S}^1_2$ due to the symmetric property.
		We note that for situation~iv), it still has the possibility to make $\phi_i(t)$ leave $\mathcal{S}$.
		Hence, we consider a robust control scheme to avoid the last two situations.
		More specifically, our objective is to maximize the possibility of the trajectory belonging to the first three situations.


Denote the ratio of $\dot{\psi}_i$ and $\dot\rho_i$ as $\beta_i$, i.e., $\beta_i=\frac{\dot\psi_i}{\dot\rho_i}$.
We note that a smaller $\beta_i$ can make $\phi_i\in\mathcal{S}^1_2$ have a higher possibility of belonging to the first three situations.
This is because $\beta_i$ determines the tangent to the state trajectory, and a smaller $\beta_i$ corresponds to a steeper slope towards the $\rho$-axis, which maximizes the possibility of the trajectory entering $\mathcal{S}_1\cup\mathcal{S}^2_2\cup\mathcal{S}^2_4$. 
%
%
%
%
In this way, the control problem in $\mathcal{S}^1_2$ is formulated as
	$$
	\begin{aligned}
	\mathrm{minimize}~&\frac{\omega_i}{v_i\sin\psi_i}-\frac{\kappa(p_i)\cot\psi_i}{1-\kappa(p_i)},\\
	\mathrm{s.t.}~~\phi_i(t)&\in\mathcal{S}^1_2,~\text{and~\eqref{eq_constraint} holds.}\\
	\end{aligned}
	$$
	and the solution is
	\begin{equation}
	v_i=v_{\min},\quad
	\omega_i=-\omega_{\max}.
	\label{eq_law21}
	\end{equation}
	Similarly, the control law in $\mathcal{S}^3_2$ can be derived as
	\begin{equation}\label{eq_law23}
	v_i=v_{\min},\quad
	\omega_i=\omega_{\max}.
	\end{equation}
	
	Then, we provide a sufficient condition for states in $\mathcal{S}^1_2$ entering $\mathcal{S}_1$ finally.

	\begin{theorem}[Dynamics in $\mathcal{S}^1_2$]
		%
		Consider the system described by state equations~\eqref{eq16} and~\eqref{eq17}, where the state variables are denoted as $\tilde{\phi}_i=(\tilde{\rho}_i,\tilde{\psi}_i)$.
		If $\tilde{\phi}_i(t_0)=\phi_i(t_0)\in\mathcal{S}^1_2$, and the state trajectory of $\tilde{\phi}_i$  has an intersection with the $\tilde{\rho}$-axis, denoted as $(R^*,0)$, where $R^*\leq R_2<\frac{1}{\kappa_0}-\frac{v_{\min}}{\omega_{\max}}$,
		then for any $\phi_i(t_0)\in\mathcal{S}^1_2$, by applying control law~\eqref{eq_law21}, $\phi_i$ will get into $\mathcal{S}_1\cup\mathcal{S}^2_2\cup\mathcal{S}^4_2$ in a finite time.
		\begin{equation}
		\begin{cases}
		\dot{\tilde{\rho}}_i=v_{\min}\sin\tilde{\psi}_i,\\
		\dot{\tilde{\psi}}_i=-\omega_{\max}-\frac{\kappa_0v_{\min}\cos\tilde{\psi}_i}{1-\kappa_0\tilde{\rho}_i},
		\end{cases} \pi/2\leq\tilde{\psi}_i<\pi
		\label{eq16}
		\end{equation}
		\begin{equation}
		\begin{cases}
		\dot{\tilde{\rho}}_i=v_{\min}\sin\tilde{\psi}_i,\\
		\dot{\tilde{\psi}}_i=-\omega_{\max}+\frac{\kappa_0v_{\min}\cos\tilde{\psi}_i}{1+\kappa_0\tilde{\rho}_i},
		\end{cases} 0\leq\tilde{\psi}_i<\pi/2
		\label{eq17}
		\end{equation}
		\label{thm_S21}
	\end{theorem}

	\begin{IEEEproof}
		The state trajectory of $\tilde{\phi}_i$ after $t_0$ can be described by $g(\phi_i)=0$, with the gradient vector $\bigtriangledown g(\phi_i)=(-\dot{\tilde{\psi}}_i,\dot{\tilde{\rho}}_i)$.
		Since $g(\phi_i)=0$ has an intersection with the $\tilde{\rho}$-axis at $(R^*,0)$, where $R^*\leq R_2$, then $g(\phi_i)=0$ lies in $\mathcal{S}$ when $\tilde{\psi}_i\in[0,\pi)$.
		By applying control law~\eqref{eq_law21}, $\dot{\rho}_i=v_{\min}\sin\psi_i$, $\dot{\psi}_i=-\omega_{\max}-\frac{\kappa(p_i)v_{\min}\cos\psi_i}{1-\kappa(p_i)\rho_i}$.
		Suppose $g(\phi_i(t))=0$ at time $t$, where $t\geq t_0$, and $\phi_i(t)\in\mathcal{S}_2^1$, 
		it can be verified that $f(\phi_i)\cdot\bigtriangledown g(\phi_i)\leq 0$ holds.
		With Lemma~\ref{lm_nagumo}, we conclude that the state trajectory of $\phi_i$ is bounded by $g(\phi_i)=0$ when $\phi_i\in\mathcal{S}^1_2$, i.e., $\phi_i$ will not leave $\mathcal{S}$.
		Let $\alpha_1=\omega_{\max}-\frac{\kappa_0v_{\min}}{1-\kappa_0R_2}$, and we have $\alpha_1>0$.
		By applying control law~\eqref{eq_law21}, $\dot{\psi}_i\leq-\alpha_1<0$ when $\phi_i\in\mathcal{S}^1_2$. Thus there exists a finite time $t_1\leq t_0+\frac{\pi}{\alpha_1}$, such that $\phi_i(t_1)\in\mathcal{S}_1\cup\mathcal{S}^2_2\cup\mathcal{S}^4_2$.
	\end{IEEEproof}

	For states in $\mathcal{S}^3_2$, we have a similar result.
	
	\begin{theorem}[Dynamics in $\mathcal{S}^3_2$]
		%
		Consider the system described by state equations~\eqref{eq19} and~\eqref{eq20}, where the state variables are denoted as $\tilde{\phi}_i=(\tilde{\rho}_i,\tilde{\psi}_i)$.
		If $\tilde{\phi}_i(t_0)=\phi_i(t_0)\in\mathcal{S}^3_2$, and the state trajectory of $\tilde{\phi}_i$  has an intersection with the $\tilde{\rho}$-axis, denoted as $(-R^*,0)$, where $R^*\leq R_2<\frac{1}{\kappa_0}-\frac{v_{\min}}{\omega_{\max}}$.
		then for any $\phi_i(t_0)\in\mathcal{S}^3_2$, by applying control law~\eqref{eq_law23}, $\phi_i$ will get into $\mathcal{S}_1\cup\mathcal{S}^2_2\cup\mathcal{S}^4_2$ in a finite time.
		\begin{equation}
		\begin{cases}
		\dot{\tilde{\rho}}_i=v_{\min}\sin\tilde{\psi}_i,\\
		\dot{\psi_i}=\omega_{\max}+\frac{\kappa_0v_{\min}\cos\tilde{\psi}_i}{1+\kappa_0\tilde{\rho}_i},
		\end{cases} -\pi\leq\tilde{\psi}_i<-\pi/2
		\label{eq19}
		\end{equation}
		\begin{equation}
		\begin{cases}
		\dot{\tilde{\rho}}_i=v_{\min}\sin\tilde{\psi}_i,\\
		\dot{\psi_i}=\omega_{\max}-\frac{\kappa_0v_{\min}\cos\tilde{\psi}_i}{1-\kappa_0\tilde{\rho}_i},
		\end{cases} -\pi/2\leq\tilde{\psi}_i\leq0
		\label{eq20}
		\end{equation}
		\label{thm_S23}
	\end{theorem}

	After $\phi_i$ enters $\mathcal{S}^2_2\cup\mathcal{S}^4_2$, it is certain that following our proposed control law in these two subsets (see Section~\ref{sec:Near Time Optimal Control Law}), $\phi_i$ can finally enter $\mathcal{S}_1$.
	We have the following theorem to conclude the stability and convergence of the overall closed-loop system.
	
	\begin{theorem}\label{controllervoerall}
		Consider a fleet of fixed-wing UAVs following a $\mathcal{C}^2$-smooth path under Assumption~\ref{asp_path}, with the path following error equation described by~\eqref{eq_patherror} with constraints~\eqref{eq_constraint}. If $R_2< \frac{1}{\kappa_0}-\frac{v_{\min}}{\omega_{\max}}$, and each UAV is in one of the following cases:
		\begin{itemize}
			\item \emph{Case 1}: $\phi_i(t_0)\in\mathcal{S}_1$;
			\item \emph{Case 2}: $\phi_i(t_0)\in\mathcal{S}_2^2\cup\mathcal{S}_2^4$;
			\item \emph{Case 3}: $\phi_i(t_0)\in\mathcal{S}_2^1$, and the state trajectory of $\tilde{\phi}_i$ following~\eqref{eq16} and~\eqref{eq17} has an intersection with the $\tilde{\rho}$-axis at $(R^*,0)$, where $R^*\leq R_2$, when $\tilde{\phi}_i(t_0)=\phi_i(t_0)$;
			\item \emph{Case 4}: $\phi_i(t_0)\in\mathcal{S}_2^3$, and the state trajectory of $\tilde{\phi}_i$ following~\eqref{eq19} and~\eqref{eq20} has an intersection with the $\tilde{\rho}$-axis at $(-R^*,0)$, where $R^*\leq R_2$, when $\tilde{\phi}_i(t_0)=\phi_i(t_0)$;
		\end{itemize}
		then by executing Algorithm~\ref{alg} when $\phi_i(t)\in\mathcal{S}_1$, control law~\eqref{eq_law24norm} and~\eqref{eq_law24abnorm} when $\phi_i(t)\in\mathcal{S}_2^4$, control law~\eqref{eq_law22norm} and~\eqref{eq_law22abnorm} when $\phi_i(t)\in\mathcal{S}_2^2$, control law~\eqref{eq_law21} when $\phi_i(t)\in\mathcal{S}_2^1$, and control law~\eqref{eq_law23} when $\phi_i(t)\in\mathcal{S}_2^3$, we will finally get
		$\lim\limits_{t\rightarrow\infty}\phi_i=\mathbf{0},~\lim\limits_{t\rightarrow\infty}\zeta_i=L,~\forall~i$.
	\end{theorem}

		Note that the proposed controller in Theorem \ref{controllervoerall} for the overall sysem is hybrid, as the controller	 is not continuous at the boundary of the subsets. For example, when $\phi_i$ enters $\mathcal{S}_1$ from $\mathcal{S}_4^2$ at time $t_a$, and $\psi_i(t_a)\geq -a+\epsilon_0$, then according to~\eqref{eq_law24norm}, $\lim\limits_{t\to t_a^-}v_i=v_{\max}$, and $\lim\limits_{t\to t_a^-}\omega_i=-\omega_{\max}$. However, it does not necessarily return $v_i=v_{\max}$ and $\omega_i=\omega_{\max}$ with Algorithm~\ref{alg} at time $t_a$. 
		Now, it is the position to analyse the stability of Theorem~\ref{controllervoerall}.

	\begin{IEEEproof}
		The convergence of the closed-loop system can be concluded by using the similar technique with Theorem~3.5 in~\cite{YingLan}.
		In each of the last three cases, $\phi_i$ will enter the coordination set, i.e., satisfies the condition in \emph{Case 1}, within a finite time:
		\begin{itemize}
			\item for \emph{Case 2}, by Theorem~\ref{thm_S2224}, there exists a time $t_1\geq t_0$ such that $\phi_i(t_1) \in \mathcal{S}_1$;
			\item for \emph{Case 3} and \emph{Case 4}, by Theorem~\ref{thm_S21} and Theorem~\ref{thm_S23}, there exists a time $t_2\geq t_0$ such that $\phi_i(t_2) \in \mathcal{S}_1\cup\mathcal{S}_2^2\cup\mathcal{S}_4^2$; if $\phi_i(t_2) \in \mathcal{S}_2^2\cup\mathcal{S}_4^2$, by Theorem~\ref{thm_S2224}, there exists a time $t_3\geq t_2\geq t_0$ such that $\phi_i(t_3) \in \mathcal{S}_1$;				
		\end{itemize}
		while for \emph{Case~1}, by Theorem~\ref{thm_invariant}, $\phi_i$ remains in $\mathcal{S}_1$ thereafter.
		Consequently, there exists a time $t^*\geq t_0$ such that $\phi_i(t^*) \in \mathcal{S}_1$ holds for all $i$.		
		Thus, following Theorem~\ref{thm_pathfollow}, we get $\lim\limits_{t\rightarrow\infty}\phi_i=\mathbf{0}$, and following Theorem~\ref{thm_coordination}, we get $\lim\limits_{t\rightarrow\infty}\zeta_i=L$.
	\end{IEEEproof}

	\begin{remark}
		We note that collisions between UAVs can be avoided if the path has no intersection points and the UAVs are inside the coordination set, since no overtaking will occur according to Theorem~\ref{thmarcdist}.
		However, when the UAVs are outside the coordination set, since they are executing the single-agent level control law, collision avoidance is not guaranteed.
		In real applications, some collision avoidance algorithms such as~\cite{YangJianJGCD} can be employed.
	\end{remark}
	\section{Simulation Results\label{sec:simulation}}
	

	In this section, simulations are given to corroborate the effectiveness of our control strategy for coordinated path following.
	The simulation consists of two parts: the simulation with MATLAB, and the Hardware-In-the-Loop (HIL) simulation with the X-Plane simulator.
	\subsection{MATLAB Simulation\label{sec:matlab}}
	Firstly, we validate the algorithm with a typical path following problem, the cyclic pursuit on a circle. The control objective is to distribute all the UAVs uniformly on a circle. The UAVs are with the constraints $v_{\min}=10\mathrm{m}/\mathrm{s}$, $v_{\max}=25\mathrm{m}/\mathrm{s}$, and $\omega_{\max}=0.2\mathrm{rad}/\mathrm{s}$. The desired path is a circle centered at $(0,0)$, with the radius of $r=1000\rm{m}$. We are employing $n=6$ UAVs in the simulation, then the desired arc distance is $L=2\pi r/n=1000\pi/3\rm{m}$. We take $\kappa_0=0.002$, then the optimized parameters for the coordination set are $a=0.6303$, $R_1=122.1297$.
		We take control parameters $k_1=1$, $k_2=R_1/a+1$, $k_3=1$, $\epsilon_0=0.05$.
		The initial positions and the orientations of the six UAVs are $(600,0,0.6\pi)^T$, $(200,580,-\pi)^T$, $(650,-160,0.3\pi)^T$, $(1100,0,-0.25\pi)^T$, $(-1100,-80,0.75\pi)^T$, and $(-200,1000,-0.25\pi)^T$. $\chi(\zeta_i)$ is defined as
	$$
	\chi(\zeta_i)=
	\begin{cases}
	v_{\min}^r,~\text{when}~\zeta_i<L-6;\\
	0.475(\zeta_i-L+6)+v^r_{\min},~\text{when}~|\zeta_i-L|\leq6;\\
	0.95(\zeta_i-L)+v_{\min}^r,~\text{otherwise}.\\
	\end{cases}
	$$
	 where $v_{\min}^r=\frac{1}{1-\kappa_0 R_1}v_{\min}$,

	\begin{figure}[!htb]
		\centering
		\includegraphics[width=2in]{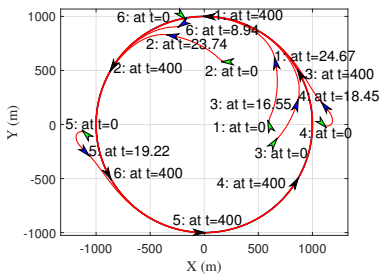}
		\caption{Trajectories of six UAVs following a circle, with the green wedges indicating the initial positions and headings for each UAV, blue ones indicating the states when the UAVs enter $\mathcal{S}_1$, black ones indicating the final states.}
		\label{fig6}
	\end{figure}

\begin{figure*}
	\centering
	\subfigure[]{\includegraphics [width=0.46\columnwidth]{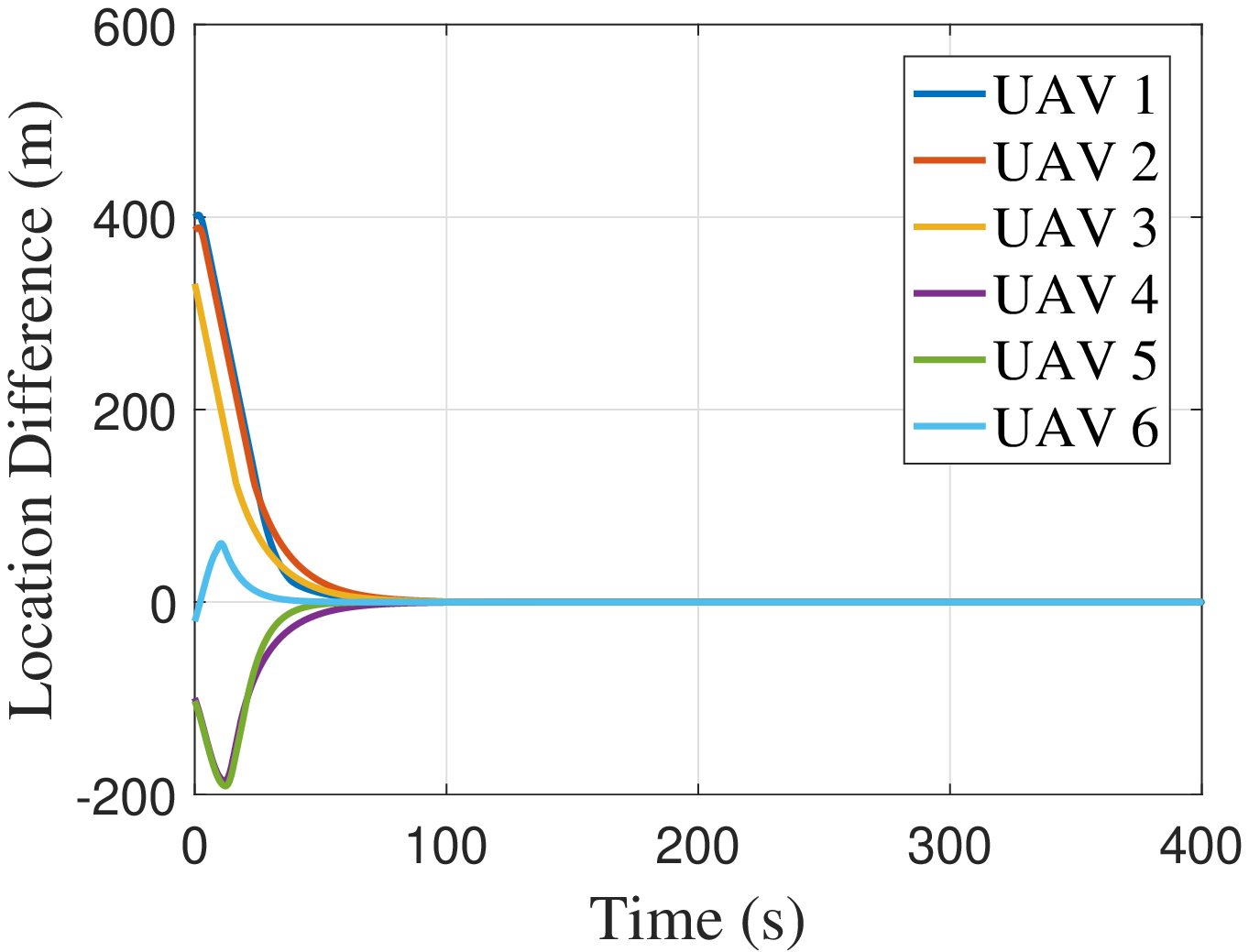}\label{fig8}}
	\hspace{0.2in}
	\subfigure[]{\includegraphics [width=0.46\columnwidth]{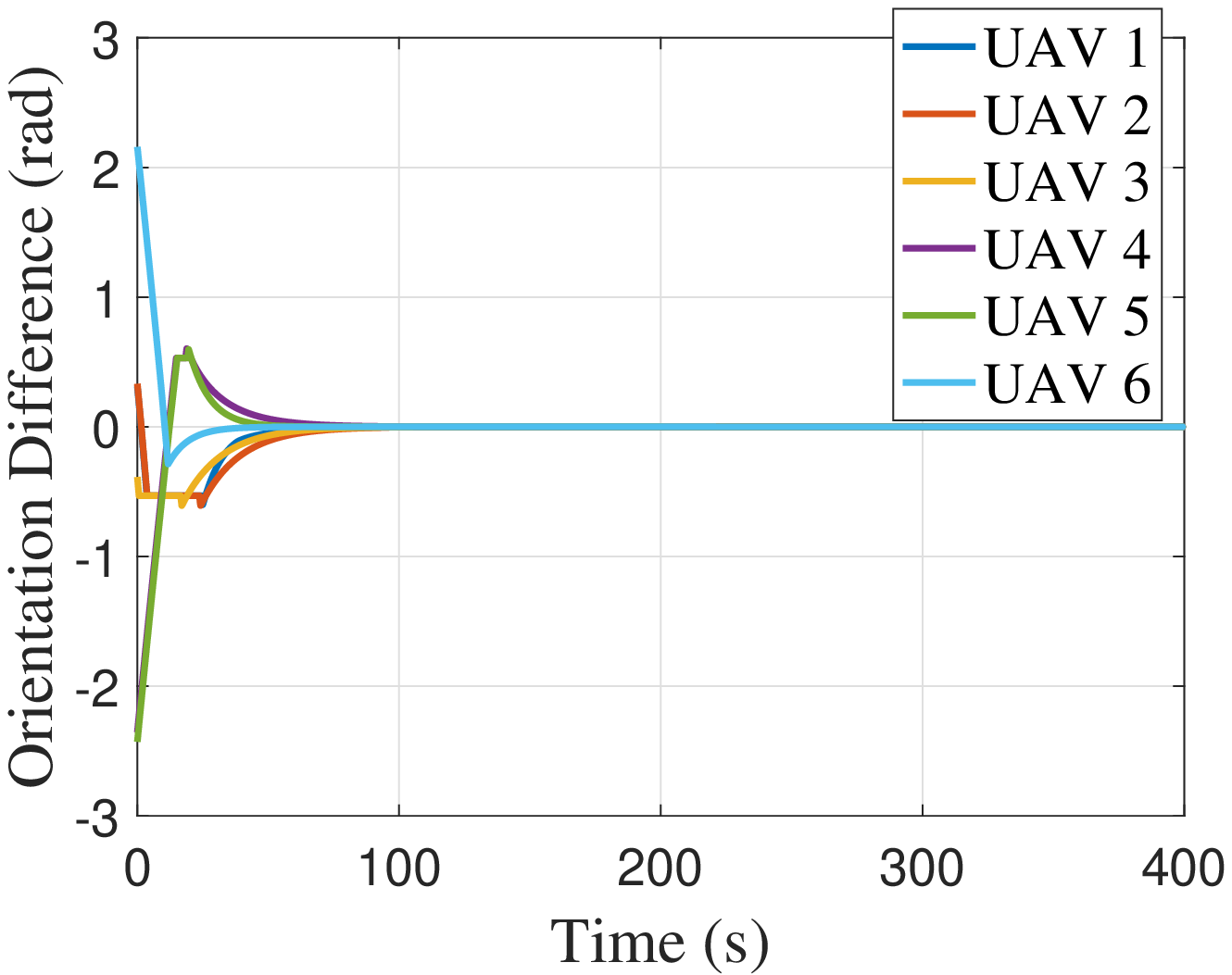}\label{fig9}}
	\hspace{0.2in}
	\subfigure[]{\includegraphics [width=0.46\columnwidth]{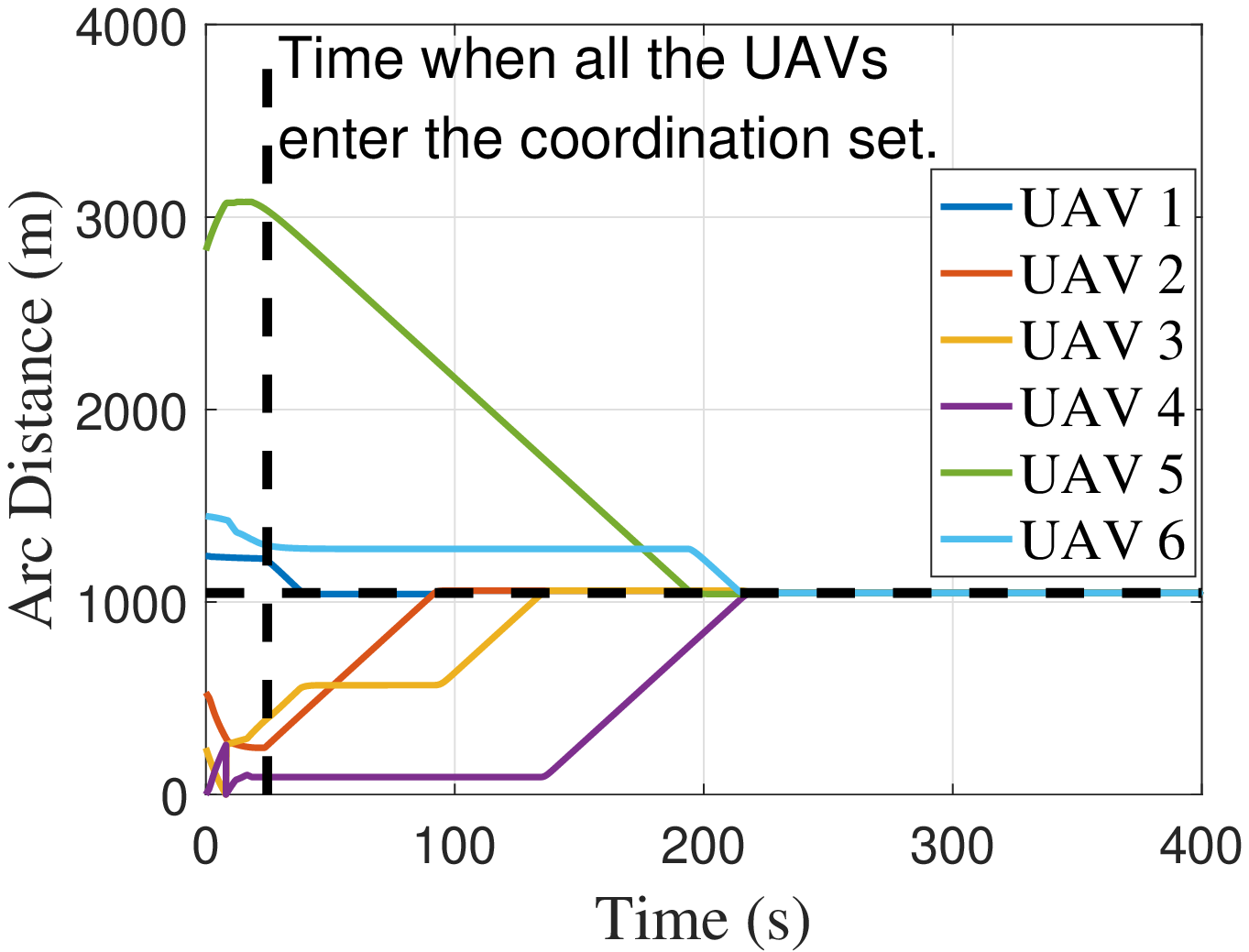}\label{fig10}}
	
	\caption{Convergence performance in MATLAB simulation:
		(a) location difference $\rho_i$ of each UAV;
		(b) orientation difference $\psi_i$ of each UAV;
		(c) arc distance $\zeta_i$ between every two adjacent UAVs.
	}\label{fig:Convergence performance MATLAB}
\end{figure*}
	
	The trajectories of the six UAVs under our hybrid control law are shown in Fig.~\ref{fig6}. The wedges in the figure not only indicate the positions of the UAVs, but also their headings.  The green wedges represent the initial positions and headings for the UAVs. Initially, $\phi_1(0),\phi_2(0),\phi_6(0)\in\mathcal{S}^1_2$, $\phi_3(0)\in\mathcal{S}^4_2$, $\phi_4(0),\phi_5(0)\in\mathcal{S}^3_2$. By executing the single-agent level control law in these subsets, $\phi_i,~i=1\ldots 6$ all enter the coordination set $\mathcal{S}_1$. 
	The blue wedges represent the positions and headings at the time when the UAVs enter $\mathcal{S}_1$. Besides, by executing the coordinated path following control law in $\mathcal{S}_1$, we find that $\phi_i(t),~i=1\ldots 6$ converge to zero, as shown in Fig.~\ref{fig8} and Fig.~\ref{fig9}, respectively. The positions and headings for the UAVs at $t=400\mathrm{s}$ are shown by the black wedges in Fig.~\ref{fig6}, and we can see that each successfully follows the path with a desired arc distance from its pre-neighbor. 


	
\begin{figure}[!htb]	
	\centering
	\includegraphics[width=2.5in]{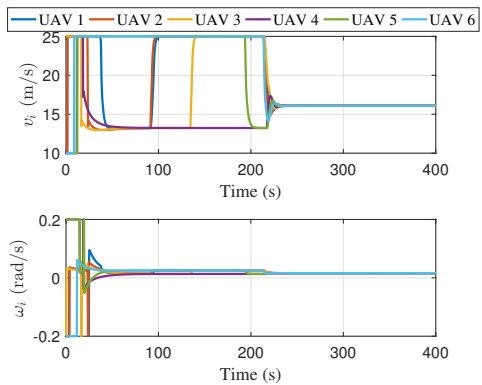}
	\caption{Control inputs of each UAV.}	
	\label{fig_control}	
\end{figure}

The arc distances between every two adjacent UAVs are shown in Fig.~\ref{fig10}, which finally converge to the desired constant $L$ after all the UAVs enter $\mathcal{S}_1$,\footnote{All the UAVs are in $\mathcal{S}_1$ after 24.67s.} meaning the coordination errors converge to 0, and all the UAVs are eventually evenly spaced with the desired distance between the adjacent UAVs, while moving along the path. We should note that there is a jump of the arc distances to 0 for UAV 3 and 4 before all the UAVs enter the coordination set $\mathcal{S}_1$, which means an overtaking occurs, and their pre-neighbors are changed. However, after all the UAVs enter $\mathcal{S}_1$, there are no such jumps of arc distances, meaning that each UAV will not change its pre-neighbor, and there are no overtakings any more.
Fig.~\ref{fig_control} shows the control inputs of each UAV. It can be observed that the control inputs are bounded but not continuous.

	\begin{figure}[!htb]
		\centering
		\includegraphics[width=2in]{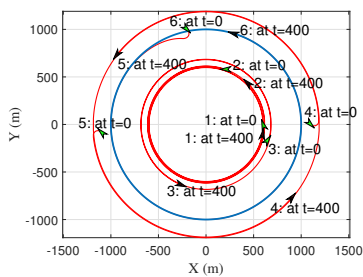}
		\caption{Trajectories of six UAVs following a circle with the method in~\cite{YingLan}, while $v_i$ and $\omega_i$ are constrained. Only UAV 6 is on the desired path at $t=400$.}
		\label{fig_fail}
	\end{figure}
		
	It should be noted that the existing methods for the coordinated path following control without considering speed constraints cannot solve our problem. The trajectories of six UAVs with speed constraints using the method in~\cite{YingLan} are shown in Fig.~\ref{fig_fail}. All the UAVs are placed at the same initial positions as in Fig.~\ref{fig6}. It can be seen that only UAV 6 finally converges to the desired path, while the other five UAVs do not, demonstrating that our proposed control law has extended the work in~\cite{YingLan} and solved the problem of coordinated path following with speed constraints.
	\subsection{HIL Simulation}
\begin{figure}[!htb]
		\centering
		\includegraphics[width=2.2in]{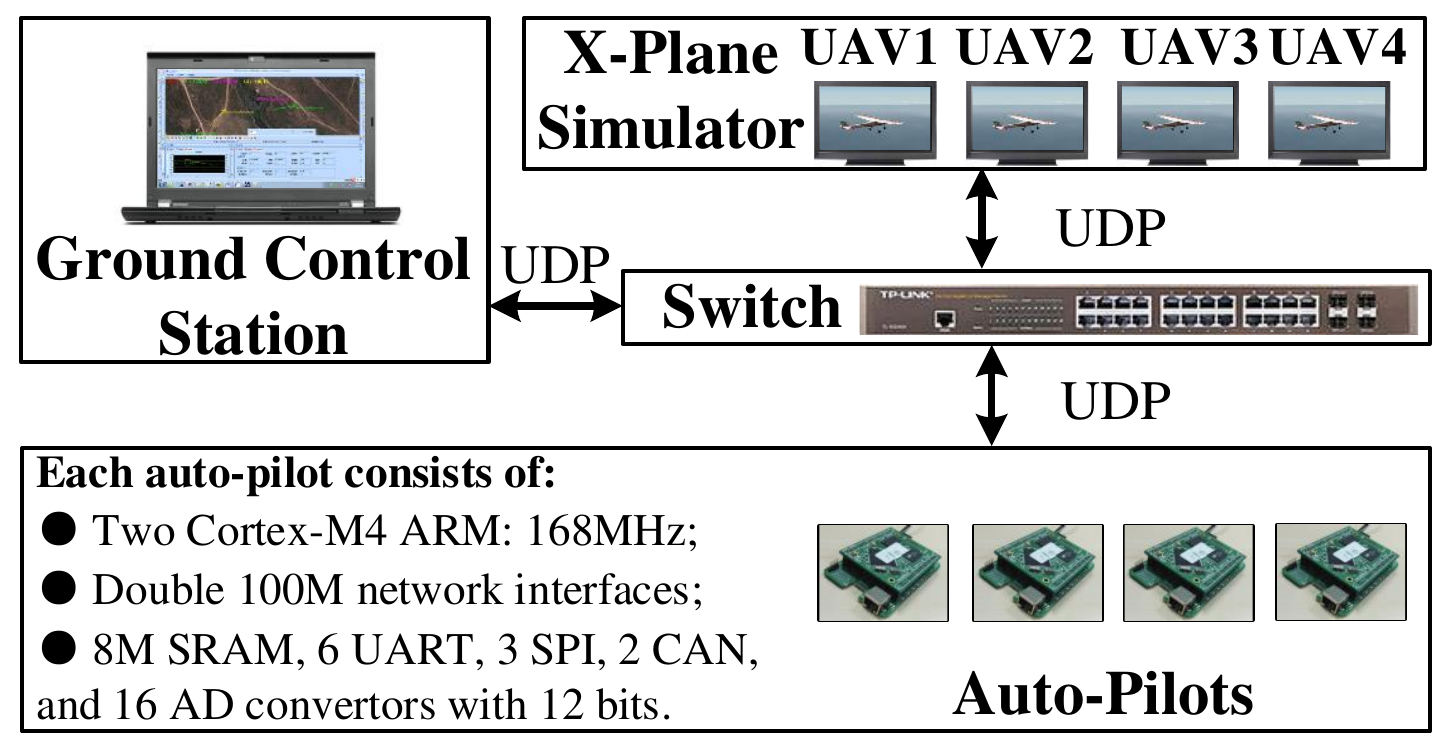}
		\caption{The hardware-in-the-loop simulation environment.}
		\label{fig_sys}
	\end{figure}

	To further validate the proposed algorithm, an HIL simulation environment is constructed, which consists of four computers running the \emph{X-Plane} 
	 flight simulator, four \emph{auto-pilots}, and a \emph{ground control station}, as shown in Fig.~\ref{fig_sys}. We use ethernet networks for the communications among the three parts. 
	The plane chosen in our HIL simulation is the \emph{Great Planes PT-60 RC plane}~\cite{Shulong2017,Shulong2018}. 
	With all these facilities, we have conducted two typical HIL simulations, i.e., cyclic pursuit and parallel path following.
	

	\subsubsection{Cyclic Pursuit}
	
\begin{figure*}
\centering
\subfigure[]{\includegraphics [width=0.46\columnwidth]{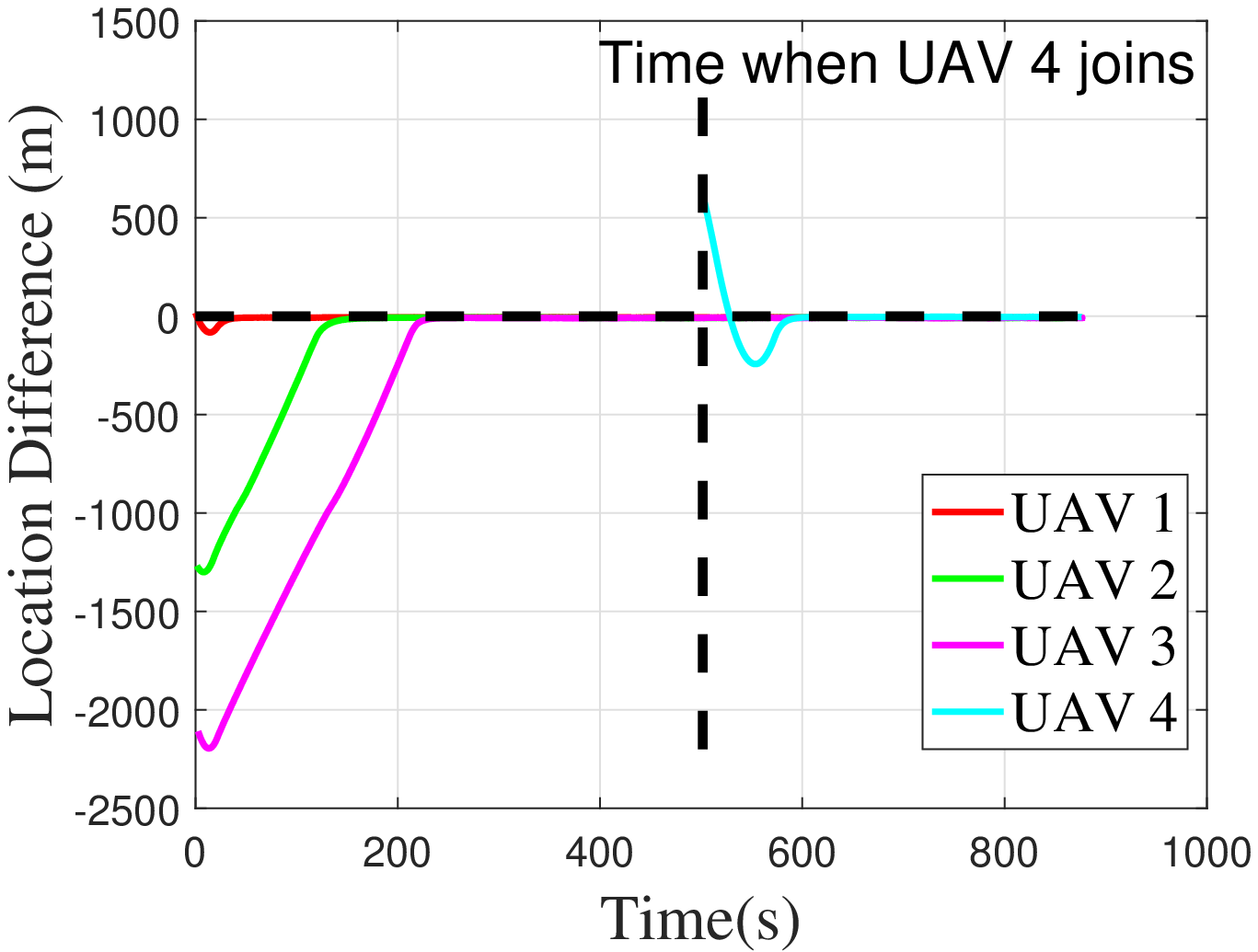}\label{fig_loccir}}
\hspace{0.2in}
\subfigure[]{\includegraphics [width=0.43\columnwidth]{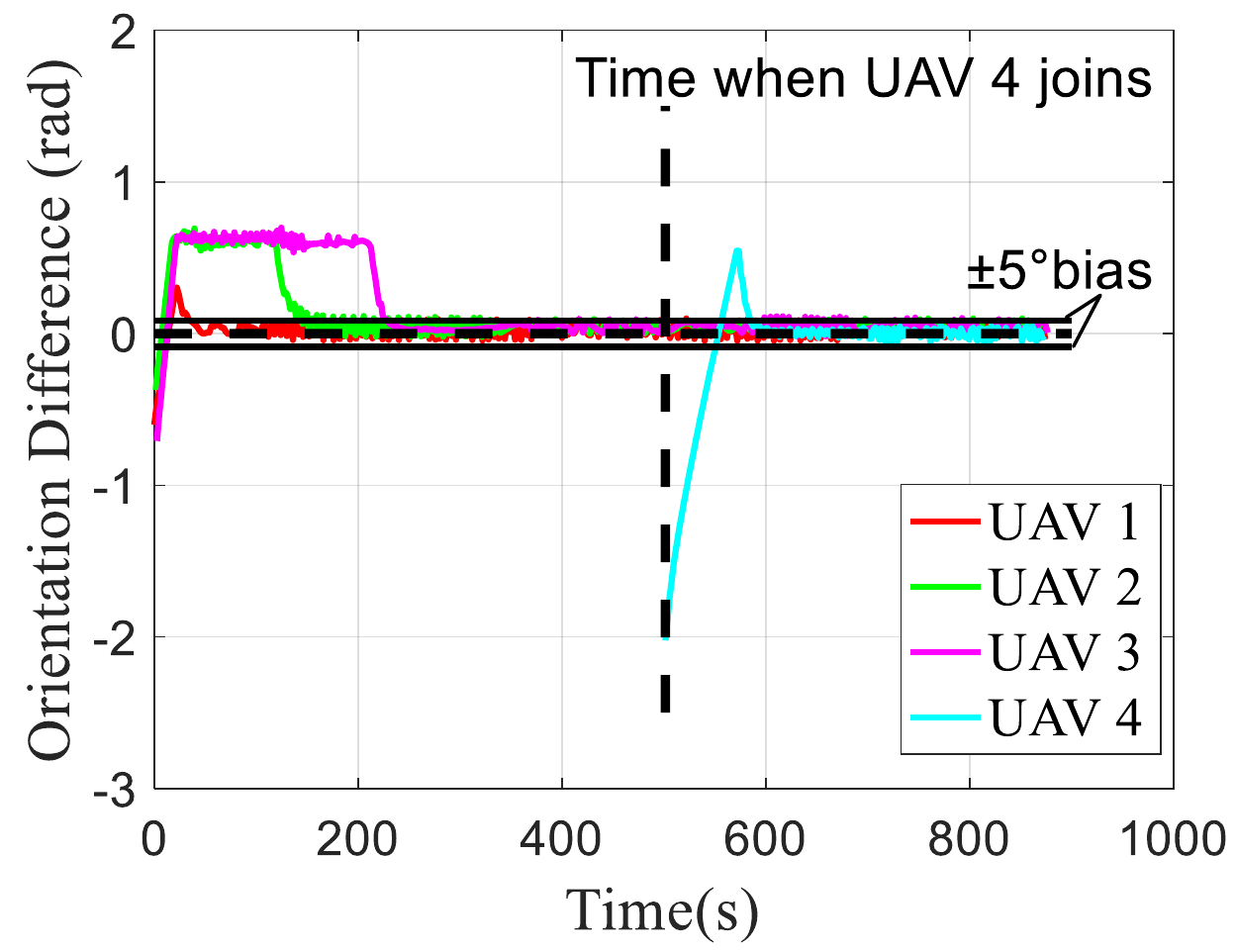}\label{fig_oricir}}
\hspace{0.2in}
\subfigure[]{\includegraphics [width=0.46\columnwidth]{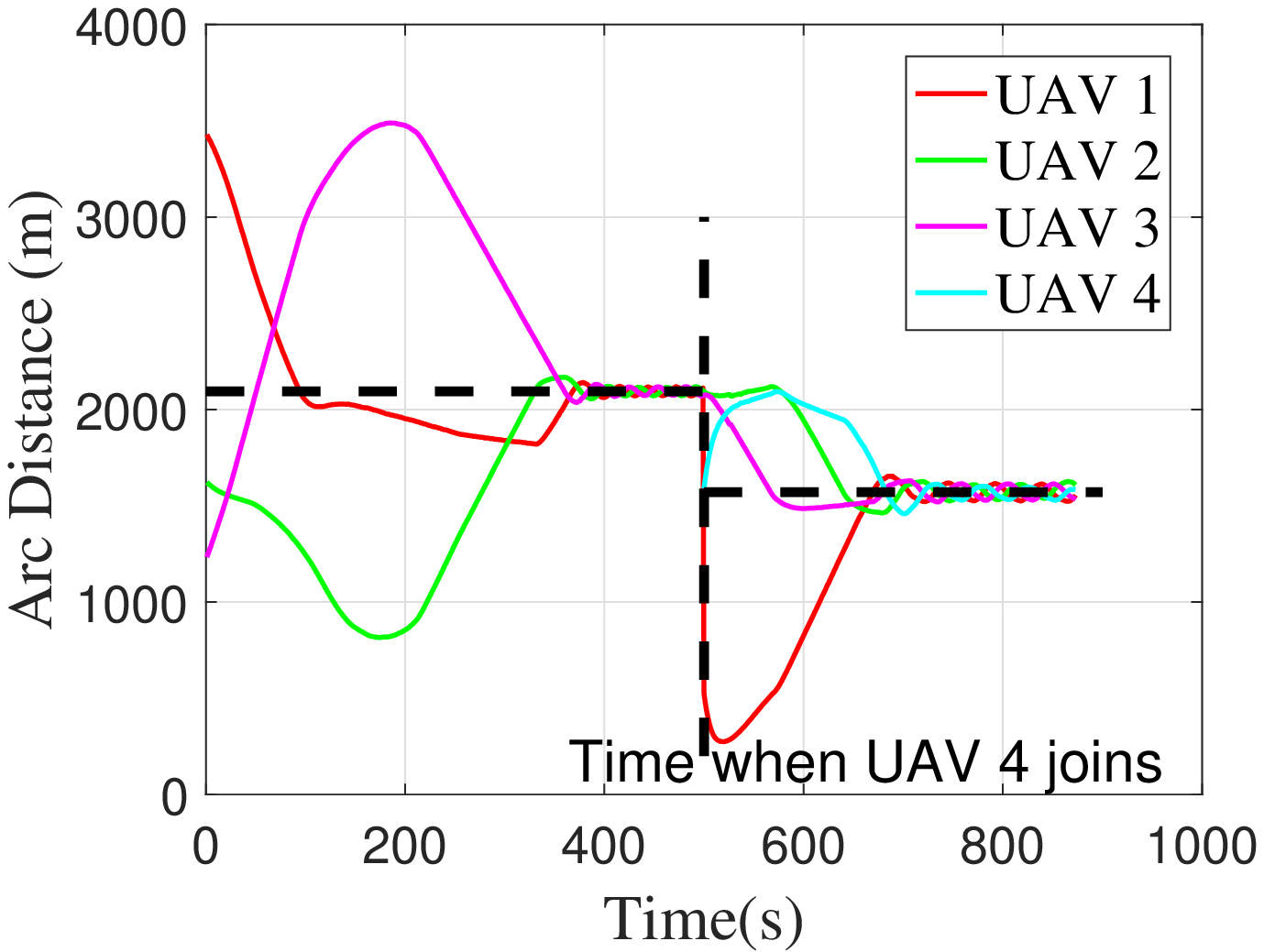}\label{fig_arccir}}

\caption{Convergence performance in HIL simulation of cyclic pursuit:
(a) location difference $\rho_i$ of each UAV;
(b) orientation difference $\psi_i$ of each UAV;
(c) arc distance $\zeta_i$ between every two UAVs.
}\label{fig:Convergence performance HIL}
\end{figure*}

	In this setting, firstly, three UAVs are executing the coordinated path following algorithm, and trying to follow the path while being distributed evenly on the orbit. After the system becomes stable, the fourth UAV joins. The orbit and the control parameters are set as the same as those in Section \ref{sec:matlab}. The location differences and orientation differences are shown in Fig.~\ref{fig_loccir} and Fig.~\ref{fig_oricir}, respectively.

%
	\begin{table*}[!htb]	
		\centering  
		\caption{Positions of waypoints.} \label{tb_wpt}						
		\begin{tabular}{cccccccc}
			\hline
			Waypoint &1 &2 &3 &4 &5 &6 &7\\\hline
			Lon/deg &113.2167 &113.2371 &113.2167 &113.1963 &113.2167 &113.2371 &113.2167 \\
			Lat/deg &28.2029 &28.2209 &28.2390 &28.2570 &28.2751 &28.2931 &28.3112 \\
			$x$/m &0 &2006.43 &4013.47 &6019.83 &8026.83 &10033.19 &12040.19\\
			$y$/m &0 &1996.54 &0 &-1997.26 &0 &1997.87 &0
			\\\hline
		\end{tabular}				
	\end{table*}	
		
	It can be seen that the location differences of the four UAVs converge to 0, and in terms of the orientation difference, though they do not strictly converge to $0$, each only has a $\pm5^\circ$ bias at the steady state. We can also see that the joining of the fourth UAV has no influence on the stability of the location difference and orientation difference of the former three UAVs. This demonstrates the scalablity of our algorithm.
	\begin{figure}[!htbp]
		\centering
		\includegraphics[height=1.8in]{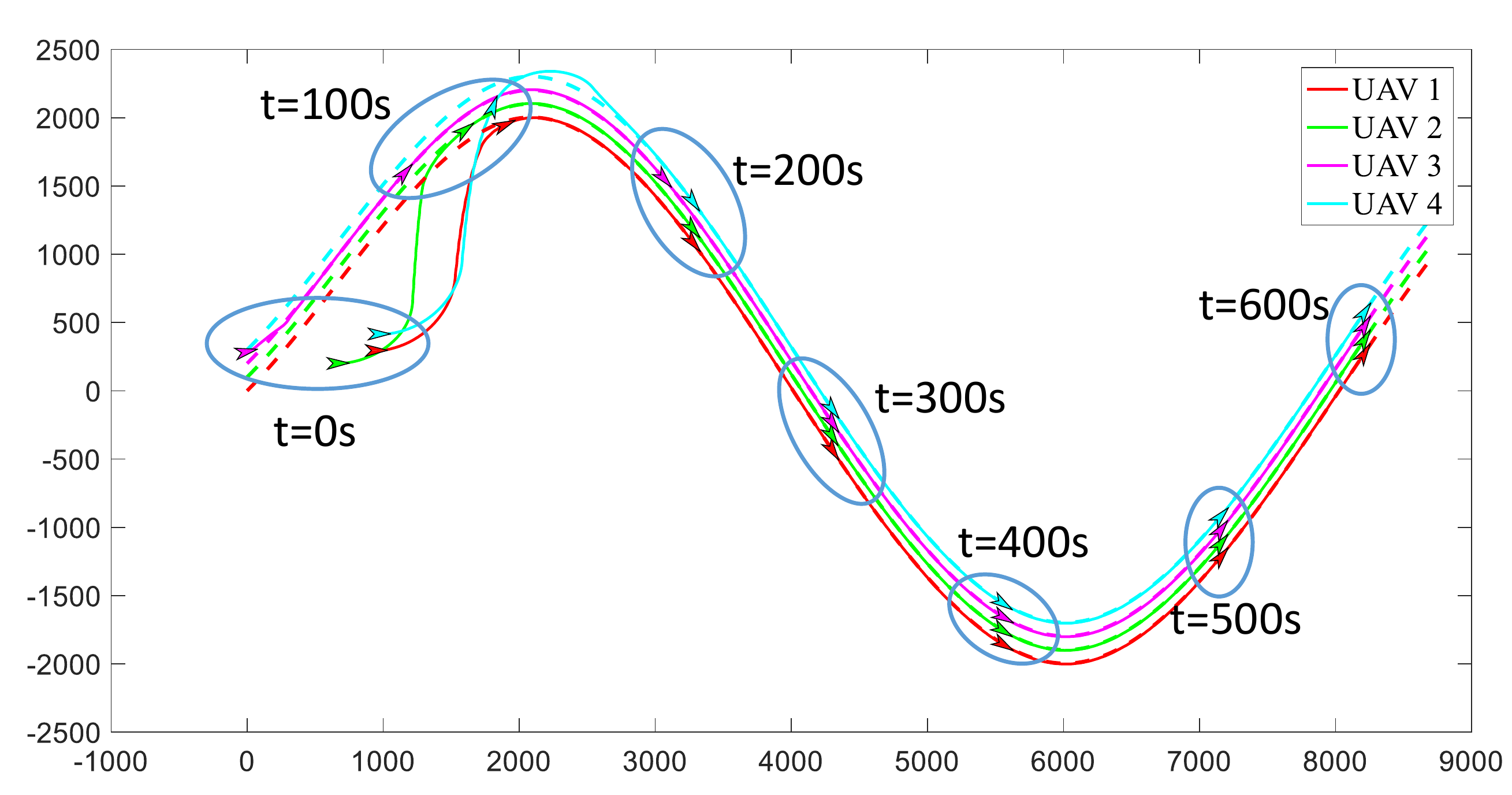}
		\caption{Parallel path Following of four UAVs, while achieving "in-line" formation pattern.}
		\label{fig_pathfollow}
	\end{figure}
	To achieve coordination, we set $p_0=(1000,0)^T$ as a reference point, and we use $l_i$ to denote the arc distance from $p_0$ to the $i$\textsuperscript{th} UAV's closest projection point $p_i$. Each UAV broadcasts its arc distance $l_i$ with respect to $p_0$ to other UAVs. After the $i$\textsuperscript{th} UAV receiving the nearby UAVs' arc distances with respect to $p_0$, it judges which UAV is its pre-neighbor, and $\zeta_i$ is calculated as $\zeta_i=l_j-l_i$, (suppose the $j$\textsuperscript{th} UAV is the $i$\textsuperscript{th} UAV's pre-neighbor). The arc distances between UAVs are shown in Fig.~\ref{fig_arccir}. We can see that the arc distances can be steered to around the desired value whether there are three or four UAVs. We can also find that there are cyclic fluctuations of arc distances. The fluctuations are caused by two main reasons. The first reason is that our control law is based on the first order system, with the acceleration period ignored in our model. The second reason is attributed to the establishment of the cyclic interaction topology in this scenario, in which each UAV has a pre-neighbor to follow. We have already shown that the proposed control law can stabilize the arc distances for the first order UAV model in Fig.~\ref{fig10}. Later we will show in another example that with a tree interaction topology, the cyclic fluctuation can be eliminated or reduced. We also note that the fewer UAVs, the smaller fluctuations would be, as shown in Fig.~\ref{fig_arccir}, where the fluctuation for 3-UAV coordination is smaller than that for 4-UAV coordination.
	\begin{figure}[!htbp]
		\centering
		\includegraphics[width=0.46\columnwidth]{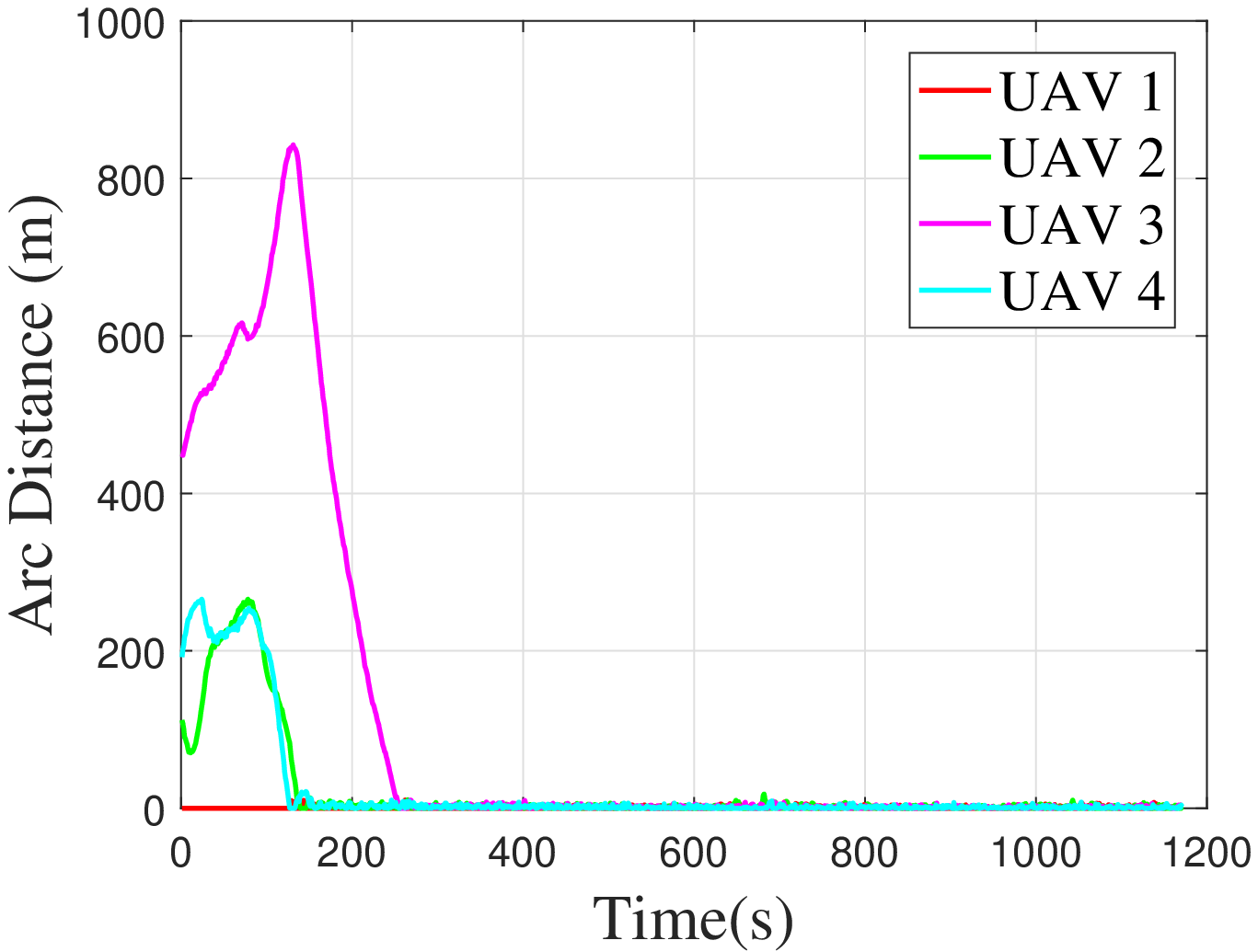}
		\caption{Arc distance $\zeta_i$ between every two UAVs in the HIL simulation of parallel path following.}
		\label{fig_arcb}
	\end{figure}

	\subsubsection{Parallel Path Following}
	 With a little bit of modification, we show our proposed hybrid control law can steer a fleet of UAVs to move on a set of parallel paths and achieve a desired ``in-line'' formation pattern. In this case, each UAV has its own target path. Still, the interaction topology is not pre-established but formed when all the UAVs enter $\mathcal{S}_1$. In order to achieve an ``in-line'' formation pattern, we set $L=0$, and $\chi(\cdot)$ is defined as $\chi(\zeta_i)=0.475\zeta_i+v^r_{\min}$.
%
	We employ a cubic B-Spline curve~\cite{Shulong2017} to obtain a continuous and non-constant curvature path. We select seven points as shown in Table~\ref{tb_wpt} to generate the cubic B-Spline for UAV 1 to follow.
	In Table~\ref{tb_wpt}, the Lon and the Lat represent the longitude and latitude, respectively.
	Besides, we establish a north-east coordinate with the origin positioned at the first waypoint of UAV 1, such that all the UAVs' states can be represented in an $xy$-plane.
	The positions of the waypoints in this new coordinate are also provided in Table~\ref{tb_wpt}.

	The simulation results of the parallel path following are shown in Fig.~\ref{fig_pathfollow}.
	The path for UAV 1 is generated by B-spline, then by moving along the $y-$direction for $100\mathrm{m}$, $200\mathrm{m}$, and $300\mathrm{m}$, we get the planned paths for UAV 2, UAV 3, and UAV 4, respectively, which are shown by the dashed curves in Fig.~\ref{fig_pathfollow}. 
	We can see that all the UAVs fly along the planned paths while achieving the desired ``in-line'' formation pattern during the flight.
		
	The arc distances between adjacent UAVs are shown in Fig.~\ref{fig_arcb}. 
	Contrary to the cyclic pursuit case with a cyclic interaction topology, the interaction topology established in this scenario is a tree, with UAV~1 as the global leader of the formation. 
	We can see that the cyclic fluctuations are eliminated in Fig.~\ref{fig_arcb}.	
	\section{Conclusion\label{sec:conclusion}}	
	In this paper, we have investigated the problem of steering a fleet of fixed-wing UAVs with speed constraints along any $\mathcal{C}^2$-smooth path with maximum curvature $\kappa_0 \leq \omega_{\max}/v_{\min}$, while achieving sequenced desired inter-UAV arc distances.
	We have proposed the hybrid control law based on the defined coordination set: for each UAV, if its path following error is within this coordination set, then the UAV follows the path in a coordination manner with its pre-neighbor; otherwise, the UAV works at the single-agent level which individually controls the path following error towards the coordination set.
	To handle the speed constraints from fixed-wing UAVs, we transform the parameter selection problem for the coordination set to an optimization problem, while satisfying the speed constraints of fixed-wing UAVs, as well as guaranteeing the convergence of both the path following error and the coordination error.
	We have also calculated the admissible set for these two errors reducing to zero when the UAVs are executing our proposed control law.
	The algorithm is validated using MATLAB and the HIL simulation, respectively, demonstrating the effectiveness of the proposed approach.
	
		The proposed approach can scale up to handle different velocity bounds for heterogeneous fixed-wing UAVs by designing different coordination sets, provided that all the UAVs have common feasible speed.
		Future work includes extending the proposed approach to the 3D case, considering communication delay, loss of communication, and wind disturbances.


	\appendices
	


	\section{Proof of Lemma~\ref{lem:Sufficient Condition under the First Principle 2}}\label{apx:Proof of Lemma lem:Sufficient Condition under the First Principle 2}
	
		When $\phi_i$ is in the first quadrant, with~\eqref{eq6}, we have
		\begin{equation}
		\label{eq_vmineq2}
		\begin{aligned}
		0&\geq v_m(a\sin\psi_i+R_1\kappa_0\cos\psi_i)-R_1\omega_{\max}+R_1\alpha\\
		&\geq v_m(a\sin\psi_i-R_1\frac{\kappa(p_i)\cos\psi_i}{1-\kappa(p_i)\rho_i})-R_1\omega_{\max}+R_1\alpha.
		\end{aligned}
		\end{equation}
		Thus~\eqref{eq_first} is derived when $v_i\in[v_{\min},v_m]$ and $\omega_i=-\omega_{\max}$.
		
		When $\phi_i$ is in the second quadrant, with~\eqref{eq_ineqprinciple2}, we have
		\begin{equation*}
		-\alpha\geq-\omega_{\max}+\frac{v_{m}\kappa_0}{1-\kappa_0R_1}\geq-\omega_{\max}-\frac{v_{m}\kappa(p_i)\cos\psi_i}{1-\kappa(p_i)\rho_i}.
		\end{equation*}		
		Thus~\eqref{eq_second} is derived when $v_i\in[v_{\min},v_m]$ and $\omega_i=-\omega_{\max}$.
		
		Inequalities~\eqref{eq_third} and~\eqref{eq_fourth} can be concluded in the same way.

	\section{Proof of Lemma~\ref{lm_second}}\label{apx:Proof of Lemma lm_second}
	
	 Denoting the speed that the $i$\textsuperscript{th} UAV moves along the path as $v^r_i$, then $v^r_i=\frac{\cos\psi_i}{1-\kappa(p_i)\rho_i}v_i$. 
    Let the $j$\textsuperscript{th} UAV be the pre-neighbor of the $i$-th UAV, and $\phi_i,\phi_j\in\mathcal{S}_1$, then $\dot{\zeta}_i=v^r_j-v^r_i$. Suppose initially $0<\zeta_i(t_0)\leq L-\delta_1$, where $0<\delta_1<L$, if~\eqref{eq7} holds, we can choose $v_i\in[v_{\min},v_m]$ and $v_j\in[v_{\min},v_{\max}]$, such that $v^r_i=\frac{1}{1-\kappa_0R_1}v_{\min}$, $v^r_j\geq\frac{1}{1-\kappa_0R_1}v_{\min}$, then $\dot{\zeta}_i\geq0$ holds for $t\geq t_0$, as a result, $\zeta_i>0$ always holds, i.e., there exists proper control law such that no overtaking occurs. 

\section{Proof of Lemma~\ref{lm_vi}}\label{apx:Proof of Lemma lm_vi}
	
		We take $\phi_i(t)\in\mathcal{S}^1_1$ as an example.
		If $\omega_i$ is not saturated in Line~\ref{ln_omegai}, i.e. $\omega_i=\omega_d$, then the value of $v_i$ will not be changed in \algoname{ReSetValue}, since the left-hand of inequality~\eqref{eq_first} becomes
		\begin{equation*}
		\begin{aligned}
		&~v_{i1}\left[a\sin\psi_i-R_1\frac{\kappa(p_i)\cos\psi_i}{1-\kappa(p_i)\rho_i}\right]+R_1\omega_i+R_1\alpha\\=&~av_{i1}\sin\psi_i-\frac{k_1R_1v_{i1}}{k_2}(k_1\rho_i+k_2\psi_i+k_3\sin\psi_i)\\
		\leq&~av_{i1}\sin\psi_i-k_1R_1v_{i1}\psi_i\stackrel{(a)}{\leq} 0.
		\end{aligned}
		\label{eq_ineqexample}
		\end{equation*}		
		where inequality~(a) is caused by $k_1R_1\geq a$ and $0<\psi_i<\pi/2$. Therefore, inequality~\eqref{eq_first} holds when $\omega_i=\omega_d$. As a result, $v_i$ will not be changed in \algoname{ReSetValue}. 
		
		Now suppose $\omega_i=\omega_{\max}$, which means $\omega_d\geq\omega_{\max}$, the left-hand of inequality~\eqref{eq_first} becomes
		$$
		\begin{aligned}
		&~v_{i1}\left[a\sin\psi_i-R_1\frac{\kappa(p_i)\cos\psi_i}{1-\kappa(p_i)\rho_i}\right]+R_1\omega_{\max}+R_1\alpha\\
		\leq&~v_{i1}\left[\frac{k_1R_1}{k_2}(k_1\rho_i+k_2\psi_i+k_3\sin\psi_i)-R_1\frac{\kappa(p_i)\cos\psi_i}{1-\kappa(p_i)\rho_i}\right]\\&~+R_1\alpha+R_1\omega_{\max}
		=R_1(-\omega_d+\omega_{\max}) \leq 0.
		\end{aligned}
		$$
		Inequality~\eqref{eq_first} holds, implying that $v_i$ will not be changed. 
		
		Finally, if $\omega_i=-\omega_{\max}$ and $v_i$ is changed in \algoname{ReSetValue}, then the returned value is
		$$
		\begin{aligned}
		v_i&=\left[a\sin\psi_i-R_1\frac{\kappa(p_i)\cos\psi_i}{1-\kappa(p_i)\rho_i}\right]^{-1}R_1(\omega_{\max}-\alpha)\\
		&\geq \frac{R_1(\omega_{\max}-\alpha)}{a\sin\psi_i+R_1\kappa_0\cos\psi_i} \stackrel{(b)}{\geq} v_m,
		\end{aligned}
		$$
		where inequality~$(b)$ follows from~\eqref{eq_vmineq2}.
		Clearly, $v_i< v_{i1}$, otherwise  inequality~\eqref{eq_first} will hold and $v_i$ does not need to be changed.
		%
		Thus, if $\phi_i\in\mathcal{S}^1_1$, and $v_i$ is changed in \algoname{ReSetValue}, then $v_m\leq v_i< v_{i1}$.		
		
		Results for the other five subsets are deduced similarly.	
	
%
%
%
%
	


	\balance
	\bibliographystyle{IEEEtran}
	\bibliography{cooppathfollow}
	
	%
	
	
	
	
	
	
	

\end{document}